\documentclass[10pt,twocolumn,letterpaper]{article}

\usepackage{cvpr}
\usepackage{times}
\usepackage{epsfig}
\usepackage{graphicx}
\usepackage{amsmath}
\usepackage{color}
\usepackage{adjustbox} 
\usepackage{subfig}
\usepackage{amssymb}
\usepackage{mathrsfs}
\usepackage{array}
\usepackage{booktabs}
\usepackage{multirow}
\usepackage{makecell}
\usepackage{float}
\usepackage{enumitem}

\def\0{{\bf 0}}
\def\1{{\bf 1}}

\def\tgen{ {\text{gen}} }
\def\tsyn{{\text{syn}}}
\def\treal{{\text{real}}}

\def\IL{{ I^L }}
\def\IH{{ I^H }}

\def\cB{{\mathcal B}}
\def\cD{{\mathcal D}}
\def\cF{{\mathcal F}}
\def\cG{{\mathcal G}}

\def\cL{{\mathcal L}}

\def\cR{{\mathcal R}}

\def\mbL{{\mathbb L}}
\def\mbH{{\mathbb H}}

\def\ie{\emph{i.e.}}
\def\eg{\emph{e.g.}}

\def\dgnote{\textcolor{black}}

\usepackage[breaklinks=true,bookmarks=false]{hyperref}

\cvprfinalcopy 

\begin{document}

\title{Learning to Zoom-in via Learning to Zoom-out: \\Real-world Super-resolution by Generating and Adapting Degradation}

\author{Dong Gong$^{*1}$,
Wei Sun$^{*2,1}$,
Qinfeng Shi$^{1}$,
Anton van den Hengel$^{1}$,
Yanning Zhang$^{2}$\thanks{* indicates equal contributions. This work was done when W. Sun was a visiting student at the University of Adelaide.}\\
$^{1}$The University of Adelaide, Australia\\
$^{2}$School of Computer Science and Engineering, Northwestern Polytechnical University, China\\
{\tt\small \!\!\!\!edgong01@gmail.com;\{wei.sun;javen.shi;anton.vandenhengel\}@adelaide.edu.au;ynzhang@nwpu.edu.cn}
}

\renewcommand\footnotemark{}

\maketitle

\begin{abstract}
Most learning-based super-resolution (SR) methods aim to recover high-resolution (HR) image from a given low-resolution (LR) image via learning on LR-HR image pairs. The SR methods learned on synthetic data do not perform well in real-world, due to the domain gap between the artificially synthesized and real LR images. Some efforts are thus taken to capture real-world image pairs. The captured LR-HR image pairs usually suffer from unavoidable misalignment, which hampers the performance of end-to-end learning, however. Here, focusing on the real-world SR, we ask a different question: since misalignment is unavoidable, can we propose a method that does not need LR-HR image pairing and alignment at all and utilize real images as they are? Hence we propose a framework to learn SR from an arbitrary set of unpaired LR and HR images and see how far a step can go in such a realistic and ``unsupervised'' setting. To do so, we firstly train a degradation generation network to generate realistic LR images and, more importantly, to capture their distribution (i.e., learning to zoom out). Instead of assuming the domain gap has been eliminated, we minimize the discrepancy between the generated data and real data while learning a degradation adaptive SR network (i.e., learning to zoom in). The proposed unpaired method achieves state-of-the-art SR results on real-world images, even in the datasets that favor the paired-learning methods more. 
\end{abstract}


\section{Introduction}
\label{sec-intro}
\begin{figure}[!t]
    \scriptsize
	\centering
    \begin{tabular}{cc}
    	\begin{adjustbox}{valign=t}
			\begin{tabular}{c}
			\subfloat{\includegraphics[width=0.304\textwidth,height=0.2\textwidth]{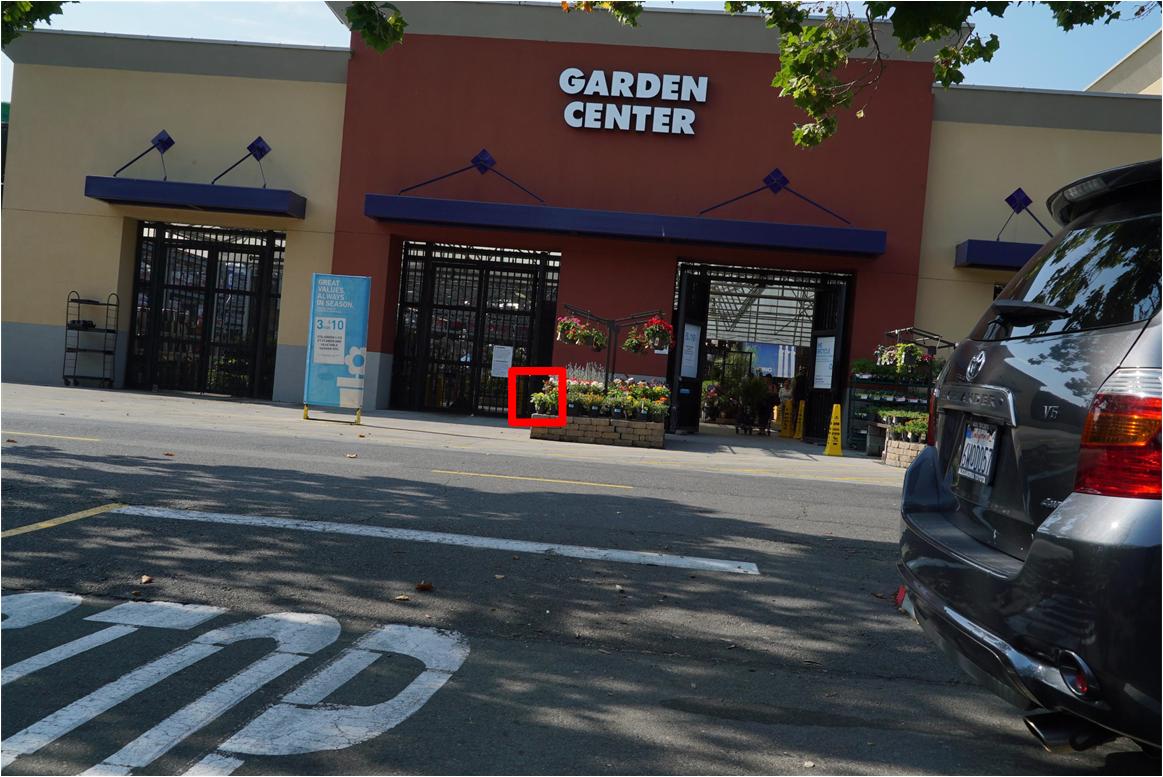}} \\
			(a) Image captured by a DSLR camera \cite{DBLP:conf/cvpr/ZhangCNK19}
			\end{tabular}
		\end{adjustbox}
		\begin{adjustbox}{valign=t}
			\begin{tabular}{c}
	        	\subfloat{\includegraphics[width=0.11\textwidth,height=0.088\textwidth]{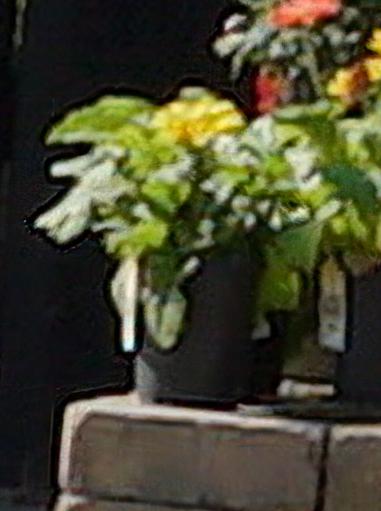}} \\
				(b) ESRGAN-Bic \cite{DBLP:conf/eccv/WangYWGLDQL18}
				\\
				\specialrule{0em}{-8pt}{0pt}
				\subfloat{\includegraphics[width=0.108\textwidth,height=0.088\textwidth]{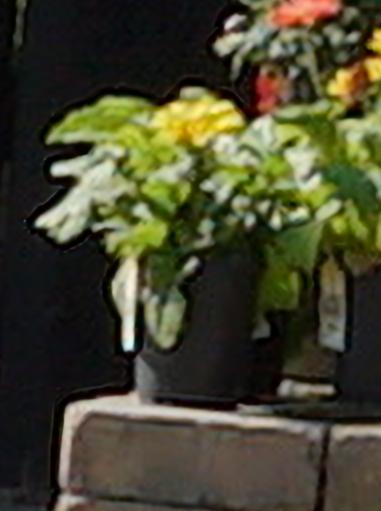}} 
			    \\
         	    (c) RCAN-Bic \cite{DBLP:conf/eccv/ZhangLLWZF18}   \\
			\end{tabular}
		\end{adjustbox} 
	\end{tabular}
	
	\begin{tabular}{c}
		\begin{adjustbox}{valign=t}
			\begin{tabular}{c@{\hspace{0.02mm}}c@{\hspace{0.5mm}}c@{\hspace{0.5mm}}c}
			\specialrule{0em}{-8pt}{0pt}
				\subfloat{\includegraphics[width=0.11\textwidth,height=0.088\textwidth]{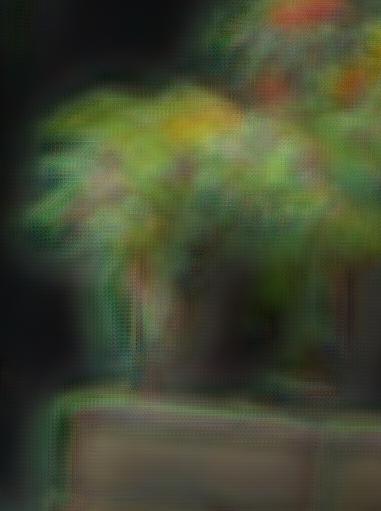}}
				&
				\subfloat{\includegraphics[width=0.11\textwidth,height=0.088\textwidth]{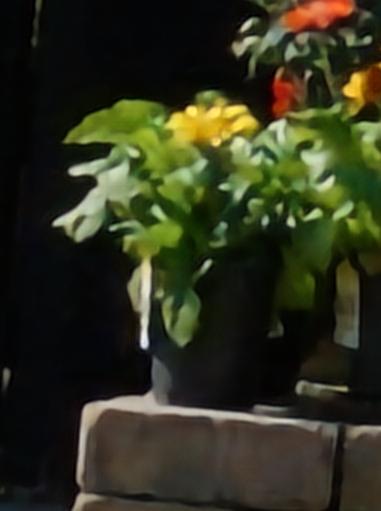}} 
				&
         		\subfloat{\includegraphics[width=0.11\textwidth,height=0.088\textwidth]{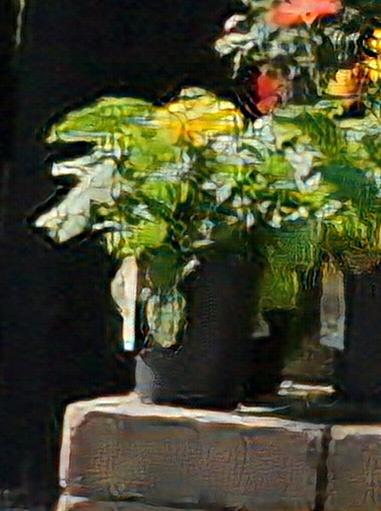}}
				&
         		\subfloat{\includegraphics[width=0.11\textwidth,height=0.088\textwidth]{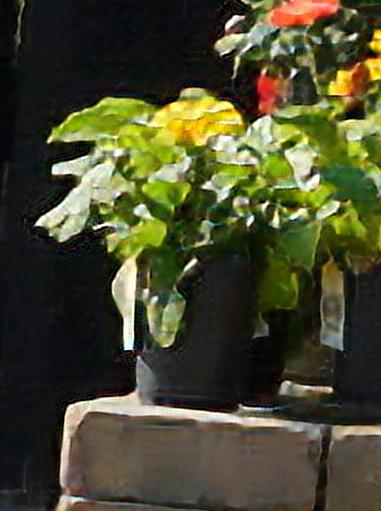}}
				\\
                 (d) RCAN-Real \cite{DBLP:conf/eccv/ZhangLLWZF18} & (e) ZoomSR \cite{DBLP:conf/cvpr/ZhangCNK19} & (f) CinCGAN \cite{DBLP:conf/cvpr/YuanLZZDL18} & (g) Ours\\
			\end{tabular}
		\end{adjustbox} 
	\end{tabular}	
    \vspace{-0.2cm}
    \caption{SR results on a real-world image captured by a digital single-lens reflex (DSLR) camera \cite{DBLP:conf/cvpr/ZhangCNK19}. ESRGAN-Bic \cite{DBLP:conf/eccv/WangYWGLDQL18} and RCAN-Bic \cite{DBLP:conf/eccv/ZhangLLWZF18} are trained based on the synthetic data with bicubic degradation. RCAN-Real \cite{DBLP:conf/eccv/ZhangLLWZF18} and ZoomSR \cite{DBLP:conf/cvpr/ZhangCNK19} are trained with nonaligned LR-HR pairs.
    The CinCGAN \cite{DBLP:conf/cvpr/YuanLZZDL18} and ours are trained under unpaired setting. Our result contains more natural details and textures suffering from less blur and artifacts. 
    }
    \vspace{-0.5cm}
    \label{fig:vis-introduction}
\end{figure}

Single image super-resolution (SR) aims to recover a high-resolution (HR) image from the corresponding low-resolution (LR) image. It is crucial and fundamental for many applications, such as mobile-phone photography and long-distance measurement. 
SR is challenging partly because recovering details from LR observations is highly ill-posed \cite{DBLP:journals/tip/YangWHM10}. 
With the recent development on deep learning, convolutional neural network (CNN) based SR methods learning from LR and HR image pairs has drawn much attention and achieved impressive results \cite{DBLP:journals/pami/DongLHT16,DBLP:conf/cvpr/LimSKNL17,DBLP:conf/eccv/ZhangLLWZF18,DBLP:conf/eccv/WangYWGLDQL18}. Due to the difficulties of capturing real LR-HR image pairs, SR CNNs \cite{DBLP:journals/corr/BrunaSL15,DBLP:conf/eccv/JohnsonAF16,DBLP:conf/iccv/0001LLG17,DBLP:conf/cvpr/ZhangTKZ018,DBLP:conf/cvpr/HarisSU18,DBLP:conf/iclr/ZhangLLZF19} are usually trained on unrealistic data synthesized by simple models, \eg, bicubic downsampling and additive Gaussian noise \cite{DBLP:conf/cvpr/ZhangZ018}.
As expected, the models trained on the bicubic synthetic data cannot generalize to real-world image SR (as shown in Figure \ref{fig:vis-introduction} (b), (c)),  due to the large \textit{distribution gap/shift} between real LR images and synthetic LR images. The degradation on the realistic images is much more complex than the simple synthetic models. 

\par
\begin{figure}[!t]
	\centering
	\scriptsize
	\scalebox{1}{
    \begin{tabular}{c@{\hspace{1mm}}c@{\hspace{1mm}}c}
         \subfloat{\includegraphics[trim=0 10 25 50, clip, width=0.15\textwidth,height=0.1\textwidth]{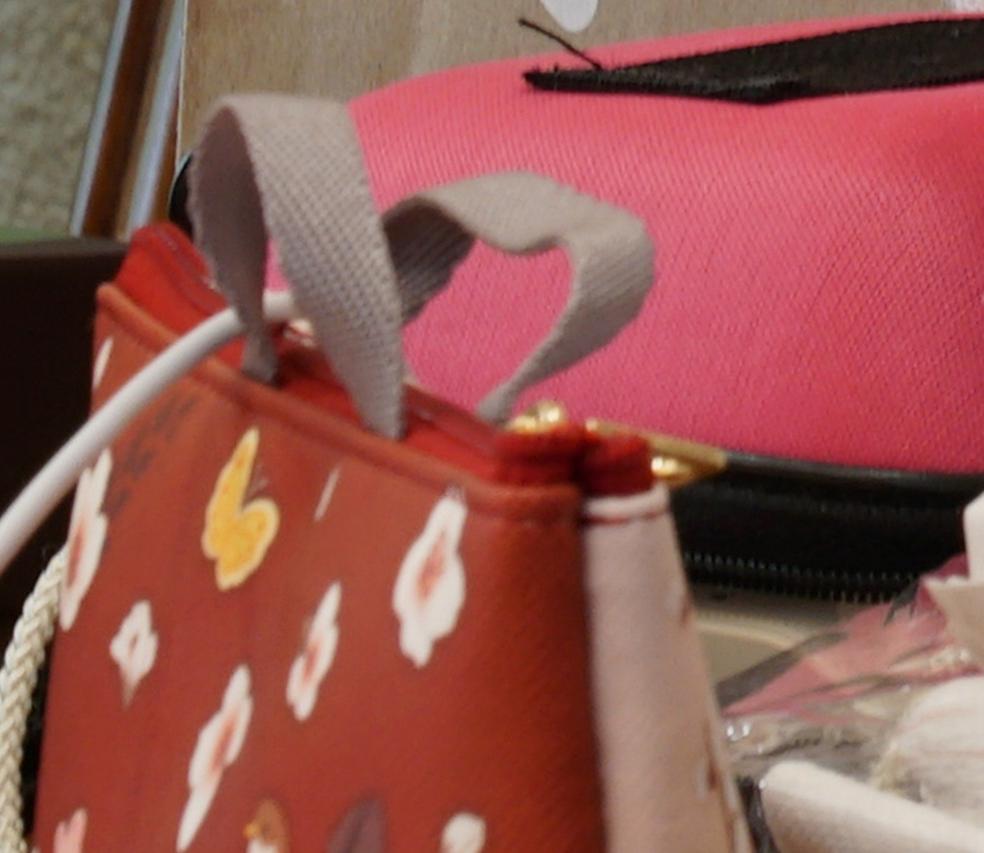}} &
         \subfloat{\includegraphics[trim=100 10 25 100,clip,width=0.15\textwidth,height=0.1\textwidth]{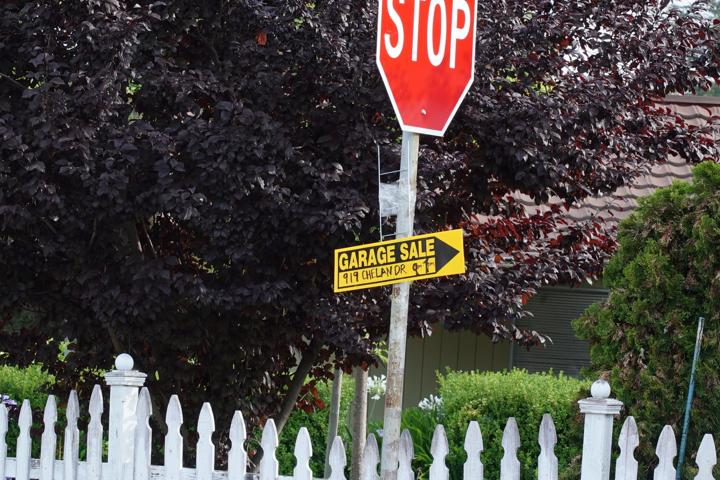}}&
         \subfloat{\includegraphics[width=0.15\textwidth,height=0.1\textwidth]{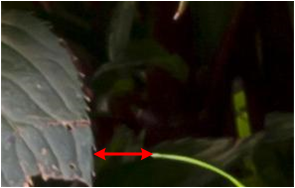}}\\
         \specialrule{0em}{-10pt}{0pt}
         \subfloat{\includegraphics[trim=0 10 25 50,clip, width=0.15\textwidth,height=0.1\textwidth]{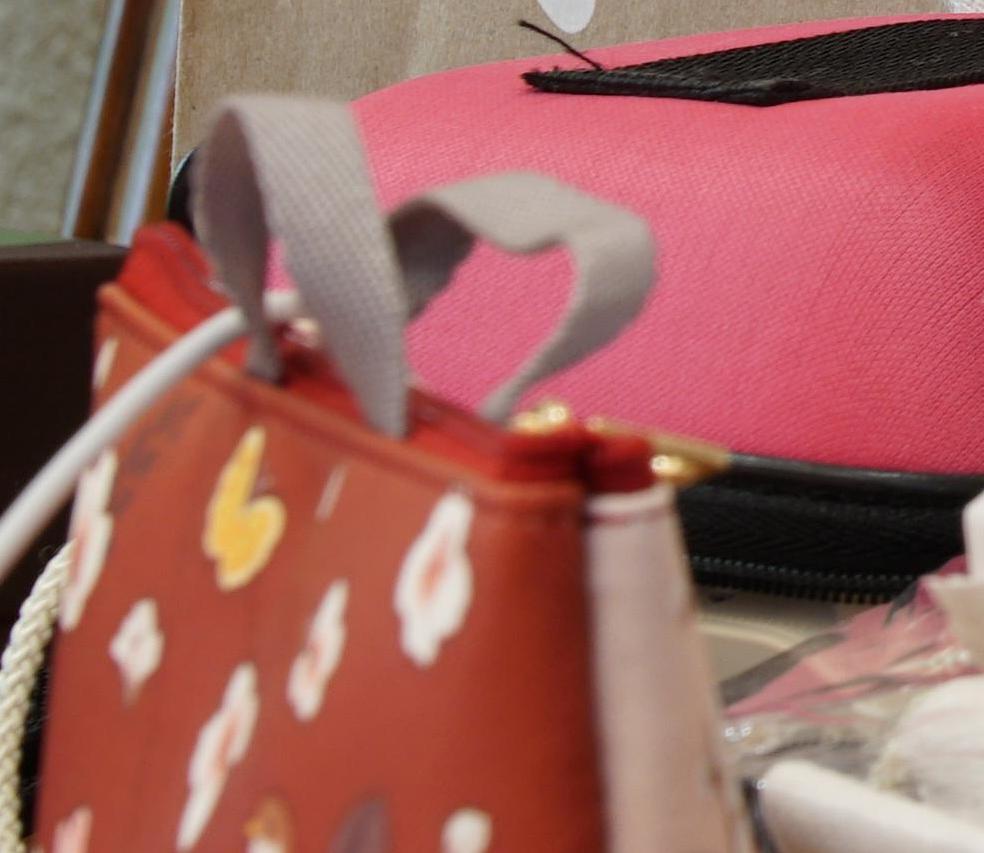}}&
         \subfloat{\includegraphics[trim=100 10 25 100,clip,width=0.15\textwidth,height=0.1\textwidth]{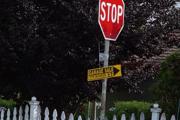}}&
         \subfloat{\includegraphics[width=0.15\textwidth,height=0.1\textwidth]{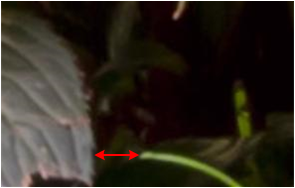}}\\ 
         
         (a) Depth-of-field & (b) Illumination & (c) Perspective
    \end{tabular}
	}
	\vspace{-0.2cm}
    \caption{Examples of misalignment between LR and HR images in real-world datasets. 
    (a) and (b) are from SR-RGB dataset \cite{DBLP:conf/cvpr/ZhangCNK19}. (c) is from RealSR dataset \cite{cai2019toward} after pre-registration. Misalignment issues are complex and extremely difficult to be removed completely.  
    }
    \vspace{-0.5cm}
    \label{fig:misalignment}
\end{figure}

Although more complicated degradation models are used for data synthesis \cite{DBLP:conf/cvpr/TimofteGWG18,DBLP:conf/cvpr/ZhangZ018}, there are still big gaps comparing to realistic data. 
For real-world SR, some recent works \cite{DBLP:conf/cvpr/ChenXTZW19,DBLP:conf/cvpr/ZhangCNK19,cai2019toward} attempted to capture realistic LR and HR images via adjusting focal length. 
Although the realistic LR images can minimize the \textit{data distribution gap}, capturing perfectly aligned LR-HR pairs is extremely hard, however. 
The \textit{misalignment} is also effected by depth-of-field, illumination, perspective, etc \cite{DBLP:conf/cvpr/ZhangCNK19} (see Figure \ref{fig:misalignment}). It cannot be eliminated via pre-processing thus making the pixel-wise training loss not really
suitable \cite{DBLP:conf/cvpr/ZhangCNK19}. 
Some efforts have been taken to register the LR and HR images by only focusing on the spatial transformation \cite{DBLP:conf/cvpr/ChenXTZW19,cai2019toward} or introduce a misalignment robust loss \cite{DBLP:conf/cvpr/ZhangCNK19}. The practicability and flexibility of these methods are very limited, however.

\par
In this paper, we propose a method towards real-world image SR based on the unpaired real LR and HR images. 
The proposed scheme requires only two unpaired image sets on LR and HR, respectively, free from image pairing supervision. 
Instead of trying to remove the \textit{LR-HR misalignment} (or capture perfectly aligned LR-HR pairs), we seek ways to bypass it by handling the \textit{domain gap} between the synthesized and realistic LR images, to achieve a more flexible and practical process. 
Despite some existing efforts in LR image generation \cite{DBLP:conf/cvpr/YuanLZZDL18,DBLP:conf/eccv/BulatYT18,DBLP:journals/corr/abs-1909-09629}, we study the problem under a much more realistic setting, in terms of both methodology and data. 
Given a set of unpaired real-world LR-HR images, we generate LR images perfectly aligned to the HR images relying on the data distribution encoded in the unpaired LR images. 
Specifically, we firstly introduce a model in the architecture of cycle-GAN \cite{DBLP:conf/iccv/ZhuPIE17} to generate realistic LR images from HR images by mimicking the real LR degradation.
The adversarial training and cycle-consistency loss are used to learn the unknown and complex characteristics of the LR images. 
Instead of assuming the generated LR images free from domain shift as \cite{DBLP:conf/cvpr/YuanLZZDL18,DBLP:conf/eccv/BulatYT18,DBLP:journals/corr/abs-1909-09629}, we propose to minimize the domain gap while training the SR network, by aligning network response (\ie, output and intermediate features) of the real and generated LR images. 
We analyze the proposed method by designing experiments on three real-world datasets \cite{DBLP:conf/cvpr/ChenXTZW19,DBLP:conf/cvpr/ZhangCNK19,cai2019toward}, which is different from previous methods that perform evaluations relying on complex but synthetic degradation. 
We ignore the pair information in the datasets and exclude the supervision signals from training. Beyond the basic setting, we also study a stricter \textit{non-overlapping} learning setting, in which the unpaired LR and HR images are enforced to have no overlapping contents. 
To summarize, the main contributions of this paper are:
\begin{itemize}[itemsep=-1.5pt,topsep=-1.5pt]
\item We propose an unpaired learning scheme for real-world image SR. 
Instead of trying to obtain well-aligned LR-HR image pairs for supervised training exhaustively, we propose to learn SR from an arbitrary set of unpaired images with LR and HR, which is a lot more general and flexible. 
Relying on the unpaired real LR and HR images, we try to generate the LR version of the real HR images (by matching the distribution of real LR images) to provide supervision for SR. 

\item We propose to reduce the domain gap between the generated and real LR images in two aspects. We firstly introduce a model to generate LR images from HR images by mimicking the real LR degradation, relying on adversarial training and cycle consistency restriction. 
Instead of assuming the generated LR images free from domain shift as previous works \cite{DBLP:conf/cvpr/YuanLZZDL18,DBLP:conf/eccv/BulatYT18,DBLP:journals/corr/abs-1909-09629}, we propose to further align the model response on the generated and real LR images while training a degradation adaptive SR network.

\item We analyze and validate the proposed method on real-world datasets, under a more realistic setting of previous synthetic degradation-based evaluation.
Extensive experiments show that the unpaired learning methods can do comparably or even better than paired learning methods under a real-world setting, and 
the proposed unpaired method achieves state-of-the-art SR results on real-world images, even in the datasets that favor the paired-learning methods more.
\end{itemize}

\section{Related Work}
\noindent \textbf{Learning-based Image Super-Resolution.}
Since the pioneer work SRCNN \cite{DBLP:journals/pami/DongLHT16}, deep learning methods have brought significant improvements in image SR. Inspired by the architecture of ResNet \cite{DBLP:conf/cvpr/HeZRS16}, EDSR \cite{DBLP:conf/cvpr/LimSKNL17} optimizes the structure of conventional residual blocks and improves the performance for SR tasks. RCAN \cite{DBLP:conf/eccv/ZhangLLWZF18} utilizes the channel attention mechanism to adaptively rescale channel-wise features and recovers more details. Furthermore, plenty of structures have been applied to the network for superior performance, such as dense connection \cite{DBLP:conf/iccv/0001LLG17}, contiguous memory \cite{DBLP:conf/cvpr/ZhangTKZ018}, back-projection \cite{DBLP:conf/cvpr/HarisSU18} and non-local \cite{DBLP:conf/iclr/ZhangLLZF19}. All the above mentioned methods rely on the ${\ell_1}$ or ${\ell_2}$-loss for training their networks, this generally leads to a blurry result \cite{DBLP:conf/cvpr/ZhangIESW18}. To address this problem, an adversarial loss \cite{DBLP:conf/nips/GoodfellowPMXWOCB14} and a perceptual loss \cite{DBLP:conf/eccv/JohnsonAF16} are integrated into the SRGAN \cite{DBLP:conf/cvpr/LedigTHCCAATTWS17}. Recently, to further enhance the visual quality, Wang et al. \cite{DBLP:conf/eccv/WangYWGLDQL18} propose ESRGAN, which combines a relativistic GAN \cite{DBLP:conf/iclr/Jolicoeur-Martineau19} for recovering more realistic details. 
Recent works \cite{DBLP:conf/cvpr/ChenXTZW19,DBLP:conf/cvpr/ZhangCNK19,cai2019toward} propose strategies to capture LR-HR image pairs via tuning focal length of DSLR cameras. 
In \cite{DBLP:conf/cvpr/ChenXTZW19,cai2019toward}, pre-registration is applied to reduce the misalignment between the captured LR and HR images. 
In \cite{DBLP:conf/cvpr/ZhangCNK19}, Zhang et. al. handle the misalignment via a robust contextural bilateral loss. Similar to most of the learning based image restoration works \cite{zhang2017beyond,gong2017motion,yang2018seeing}, imaging model (\eg bicubic downsampling model) is used to generate synthetic dataset for training \cite{DBLP:journals/pami/DongLHT16}. 
The influence of the misalignment cannot be avoidable or removed, due to the complex imaging process in camera.

\begin{figure}[!t]
    \small
	\centering
	\scalebox{1}{
    \begin{tabular}{c@{\hspace{0.1mm}}c@{\hspace{0.1mm}}c@{\hspace{0.1mm}}c}
         \subfloat{\includegraphics[width=0.12\textwidth,height=0.1\textwidth]{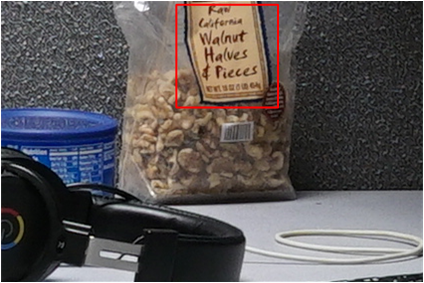}} &
         \subfloat{\includegraphics[width=0.11\textwidth,height=0.1\textwidth]{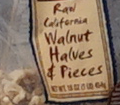}} &
         \subfloat{\includegraphics[width=0.11\textwidth,height=0.1\textwidth]{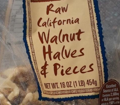}} &
         \subfloat{\includegraphics[width=0.11\textwidth,height=0.1\textwidth]{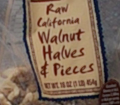}}\\    
         (a)  & (b)  & (c)  & (d)  

    \end{tabular}
	}
	\vspace{-0.2cm}
    \caption{Visual examples of different domains of LR images. (a) The original LR image from \cite{DBLP:conf/cvpr/ZhangCNK19} taken by a DSLR camera with a shorter focal length. (b) Local area from (a). (c) Bicubic synthetic image. (d) Generated image by our degradation generation network. The degradation in the realistic LR image in (b) is much more complex than the bicubic downsampling image in (c). Our generated network produces similar characteristics shown in (b).}
    \vspace{-0.5cm}
    \label{fig:degeneration}
\end{figure}

\noindent \textbf{Unpaired Super-Resolution.}
A few recent works provide a feasible solution for unpaired learning. DualGAN \cite{DBLP:conf/iccv/YiZTG17} and cycle-GAN \cite{DBLP:conf/iccv/ZhuPIE17} present an interesting network structure that contains cycle consistency on top of pix2pix \cite{DBLP:conf/cvpr/IsolaZZE17}. Both use a pair of forward and inverse generator. The forward generator maps a source domain to a target domain, while the inverse generator enforces each generated image to map back to its source domain. Inspired by these methods, Yuan et al. \cite{DBLP:conf/cvpr/YuanLZZDL18} propose a CinCGAN network, which learns to map the noise and blurry input LR image to a clean LR image space with cycle consistency losses. Then the SR network is trained by only employing indirect supervision in the LR domain. In contrast, our SR network includes direct supervision in the HR domain, resulting in better performance. In addition, we propose to further minimize the domain gap by aligning the feature distribution while training SR network. Recently, \cite{DBLP:journals/corr/abs-1909-09629} proposes an unpaired approach to learn a network that generates realistic LR images for training SR network. It assumes the generated LR images are free from domain shift/gap and trains SR model directly on the generated images.

\section{The Proposed Method}
\subsection{Problem Formulation}
The goal of SR is to increase the resolution of a given LR image $I^L$ and obtain the corresponding HR image $I^H$. SR is an inverse process of the degeneration from $I^H$ to $\IL$. To solve the ill-posed inverse problem, learning-based methods \cite{DBLP:conf/cvpr/ChenXTZW19,DBLP:conf/cvpr/ZhangCNK19,cai2019toward} seek to learn a mapping function $\cR(\cdot)$ from a set of LR-HR image pairs, \ie, $\{(I^L_i, I^H_i)\}$, to perform SR via $\IH=\cR(\IL)$. 
Since taking real-world image pairs $\{(I^L_{\treal, i}, I^H_{\treal, i})\}$ is extremely difficult, many methods train and evaluate the models by synthesizing LR image $I^L_\tsyn$ from the HR image \cite{DBLP:journals/corr/BrunaSL15,DBLP:conf/eccv/JohnsonAF16,DBLP:conf/iccv/0001LLG17,DBLP:conf/cvpr/ZhangTKZ018,DBLP:conf/cvpr/HarisSU18,DBLP:conf/iclr/ZhangLLZF19}. The most common way to obtain $I^L_\tsyn$ is downsampling $I^H_\treal$ via bicubic degradation and adding Gaussian noise. 
However, bicubic degradation is very different from the process of camera sensor sampling, as discussed in Section \ref{sec-intro} and Figure \ref{fig:degeneration}. SR model trained with the synthetic $I^L_\tsyn$ cannot generalize to real images well, due to the domain gap. 
Although some studies try to capture $\{(I^L_{\treal, i}, I^H_{\treal, i})\}$ via adjusting camera focal length \cite{DBLP:conf/cvpr/ChenXTZW19,DBLP:conf/cvpr/ZhangCNK19,cai2019toward}, the paired $I^L_{\treal}$ and $I^H_{\treal}$ suffer from severe misalignment, even after registrations (see Figure \ref{fig:misalignment}), making pixel level supervised learning unsuitable.  
Instead of trying to align $I^L_{\treal}$ and $I^H_{\treal}$ exhaustively, we seek to bypass it via performing unpaired learning on a set of unpaired real LR and HR images $\{ \{I^L_{\treal}\}, \{I^H_{\treal}\}\}$. 
It is much easier to capture extensive unpaired LR-HR images in real world. 
Our goal is thus to learn realistic SR on the unpaired datasets. 
Although some real-world datasets provide fairly aligned LR-HR pairs under some restrictions, we would like to push the boundaries of the possibilities in such flexible ``unsupervised'' setting for real-world images.

\subsection{Overview of the Proposed Method}
Given a set of unpaired LR and HR images $\{ \{I^L_{\treal, i}\}_{i=1}^M, \{I^H_{\treal,j}\}_{j=1}^N\}$, we aim to learn a SR function $\cR(\cdot)$ that maps an observed $I^L_{\treal}$ to its HR version following the distribution defined by $\{ I^H_{\treal,j}\}_{j=1}^N$ in testing. 
Considering that the detail information is crucial for the HR image quality, instead of transferring between LR and HR domain blindly, we train SR network by generating supervision first.

\begin{figure*}[!t]
\begin{center}
    \includegraphics[width=1\textwidth,height=0.25\textwidth]{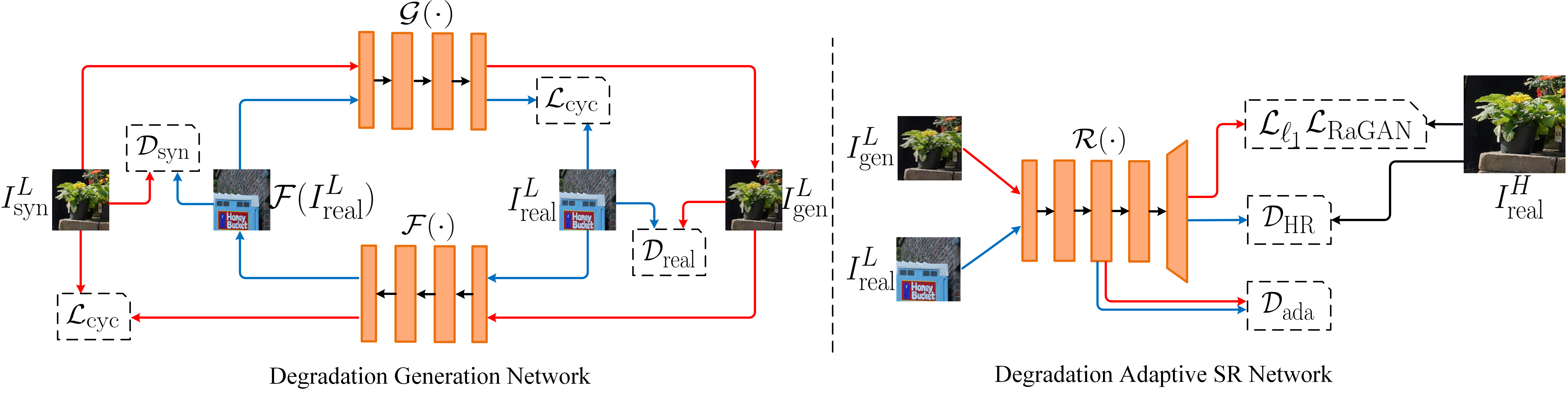}
\end{center}
\vspace{-0.6cm}
   \caption{The proposed scheme consists of two major parts -- a degradation generation network for generating LR images from real HR images and a degradation adaptive SR network for super-resolving real-world LR images. The data flow of the synthetic domain and realistic domain are denoted by red and blue lines, respectively. Firstly, we learn to generate $I_\text{gen}^L$ for $I_\text{real}^H$ by mimicking the degradation encoded in $I_\text{real}^L$. After obtaining the aligned image pair set $\{(I_\tgen^L, I_\treal^H)\}$, we train the real-world SR network $\cR(\cdot)$ under paired supervision.}
   \vspace{-0.5cm}
\label{fig:flowmap}
\end{figure*}

Firstly, we generate the realistic LR version for $\{I^H_{\treal,j}\}_{j=1}^N$, denoted as $\{I^L_{\tgen,j}\}_{j=1}^N$. Since the degradation in $I^L_{\treal}$ is unknown and complex, we learn to degrade $I^H_{\treal}$ by mimicking the degradation encoded in $\{I^L_{\treal}\}$. Specifically, we train a network $\cG(\cdot)$ to generate $I^L_\tgen$. The gap between the data distribution of $\{I^L_{\treal}\}$ and $\{I^L_{\tgen}\}$, \ie, $p(I^L_\treal)$ and $p(I^L_\tgen)$, is minimized. Since the downsampling process is controllable, we can obtain a paired LR-HR  training dataset $\{ (I^L_{\tgen}, I^H_{\treal})\}$. 

We then train a real-world SR model $\cR(\cdot)$ based on $\{ (I^L_{\tgen}, I^H_{\treal})\}$, which makes the pixel-wise training loss usable. 
Instead of assuming $\{I^L_{\tgen}\}$ can perfectly reflect the characteristics of $\{I^L_{\treal}\}$, we propose to reduce the domain gap further while training $\cR(\cdot)$, which is different from previous related SR methods \cite{DBLP:conf/cvpr/YuanLZZDL18,DBLP:conf/eccv/BulatYT18,DBLP:journals/corr/abs-1909-09629}.
During training, we apply an \emph{adaptive loss} to confuse the network response (\ie, features) of $I^L_{\tgen}$ and $I^L_{\treal}$. As a result, the SR model trained with the generated LR images can adapt to the real LR degradation insensitively. 

As shown in Figure \ref{fig:flowmap}, the proposed scheme consists of two major parts -- a \textit{degradation generation network} for generating LR images from real HR images and a \textit{degradation adaptive SR network} for super-resolving real-world LR images.

\subsection{Degradation Generation Network}
The task of degradation generation network $\cG(\cdot)$ is to obtain the LR version of $I^H_\treal$, \ie, $I^H_\tgen$, with realistic degradation. 
Given an $I^H_\treal$, we firstly downsample it as $I^L_{\tsyn}$ using bicubic operator $\cB(\cdot)$, which is aligned with $I^H_\treal$. 
The task of learning $\cG(\cdot)$ can thus be achieved by learning to transfer from domain of $\{I^L_\tsyn\}$ to $\{I^L_\treal\}$.
Although it is also an unpaired learning problem, it is much easier than transferring from $\{I^L_\tsyn\}$ to $\{I^H_\treal\}$, since it focuses only on the degradation. 

\par
We design $\cG(\cdot)$ with the architecture of cycle-GAN \cite{DBLP:conf/iccv/ZhuPIE17}. 
A reverse mapping network $\cF(\cdot)$ is used to assistant the training of $\cG(\cdot)$ by mapping the generated $\{I^L_\tgen\}$ back to $\{I^L_\tsyn\}$. 
Two adversarial discriminators \cite{DBLP:conf/cvpr/IsolaZZE17,DBLP:conf/eccv/LiW16}
$\cD_\tsyn(\cdot)$ and $\cD_\treal(\cdot)$ are involved to capture  the distribution of input synthetic LR images $\{I^L_\tsyn\}$ and  real LR images $\{I^L_\treal\}$. 
\dgnote{$\cD_\tsyn(\cdot)$ is used to distinguish the samples of $\{ \cF(I^L_\treal) \}$ from $\{ I^L_\tsyn \}$; $\cD_\treal(\cdot)$ learns to distinguish the generated samples $\{ I^L_\tgen \}$ from the realistic samples $\{ I^L_\treal \}$.}
Relying on the two discriminators, we can apply adversarial losses for matching the distribution of the generated LR images and the real LR images. 
By letting $p(I_{\tsyn}^{L})$ and $p(I_{\treal}^{L})$ denote the distribution of $\{ I_{\tsyn}^{L} \}$ and $\{ I_{\treal}^{L} \}$, respectively, we define the adversarial training objective for $\cG$ as:
\begin{equation}
\begin{aligned}
    \mathcal{L}_\text{GAN} & (\mathcal{G},\mathcal{D}_{\treal}) =  \mathbb{E}_{I_{\treal}^{L} \sim  p(I_{\treal}^{L})} [\log \mathcal{D}_{\treal}(I_{\treal}^{L})] \\
    &+ \mathbb{E}_{I_{\tsyn}^{L} \sim  p(I_{\tsyn}^{L})} [\log(1 - \mathcal{D}_{\treal}(\mathcal{G}(I_{\tsyn}^{L})))].
\end{aligned}
\label{eq_GAN}
\end{equation}
$\cG$ is optimized by minimizing the objective in Eq.\eqref{eq_GAN} against an adversarial $\cD_\treal$ that tries to maximize the loss. 
Similarly, the adversarial loss is also applied for $\cF$ via $\cD_\tsyn$. 

\par
To maintain the consistency of image contents before and after processing,  we use the cycle consistency loss \cite{DBLP:conf/iccv/ZhuPIE17} to provide additional supervision signals, which is defined as: 
\begin{equation}
\begin{aligned}
    \mathcal{L}_\text{cyc} (\cG, \cF) &= \mathbb{E}_{I_{\tsyn}^{L} \sim  p (I_{\tsyn}^{L})} [\left \| \mathcal{F}(\mathcal{G}(I_{\tsyn}^{L})) - I_{\tsyn}^{L}  \right \|_{1}] \\
    &+ \mathbb{E}_{I_{\treal}^{L} \sim  p(I_{\treal}^{L})} 
    [\left \| \mathcal{G}(\mathcal{F}(I_{\treal}^{L})) - I_{\treal}^{L}  \right \|_{1}].
\end{aligned}
\label{eq_CYC}
\end{equation}

\par
The fully objective for training $\cG(\cdot)$ is a weighted sum of the above three loss functions: 
\begin{equation}
\begin{aligned}
    \mathcal{L}_\text{LR-Gen} (\cG, & \cF, \cD_\treal, \cD_\tsyn)  = w_{1}  \mathcal{L}_\text{GAN}(\mathcal{G},\mathcal{D}_{\treal}) \\ 
    &  + w_{2}  \mathcal{L}_\text{GAN}(\mathcal{F},\mathcal{D}_{\tsyn}) + w_{3} \mathcal{L}_\text{cyc}(\cG, \cF),
\end{aligned}
\label{eq_totalloss_gan}
\end{equation}
where $w_{1}$, $w_{2}$ and $w_{3}$ are important weights. 
The degradation generation network is trained by minimizing $\mathcal{L}_\text{LR-Gen}$ w.r.t. $\cG$ and $\cF$ and maximizing w.r.t. $\cD_\treal$ and $\cD_\tsyn$. 
Relying on trained $\cG(\cdot)$, we can get the paired SR training dataset $\{(I_\tgen^L, I_\treal^H)\}$ via $I_\tgen^L=\cG(I_\tsyn^L)$ and $I^L_\tsyn=\cB(I^H_\treal)$.

\begin{figure*}[!t]
    \scriptsize
	\centering
	\scalebox{0.8}{
    \begin{tabular}{cc}
    	\begin{adjustbox}{valign=t}
			\begin{tabular}{c}
				\subfloat{\includegraphics[width=0.23\textwidth,height=0.17\textwidth]{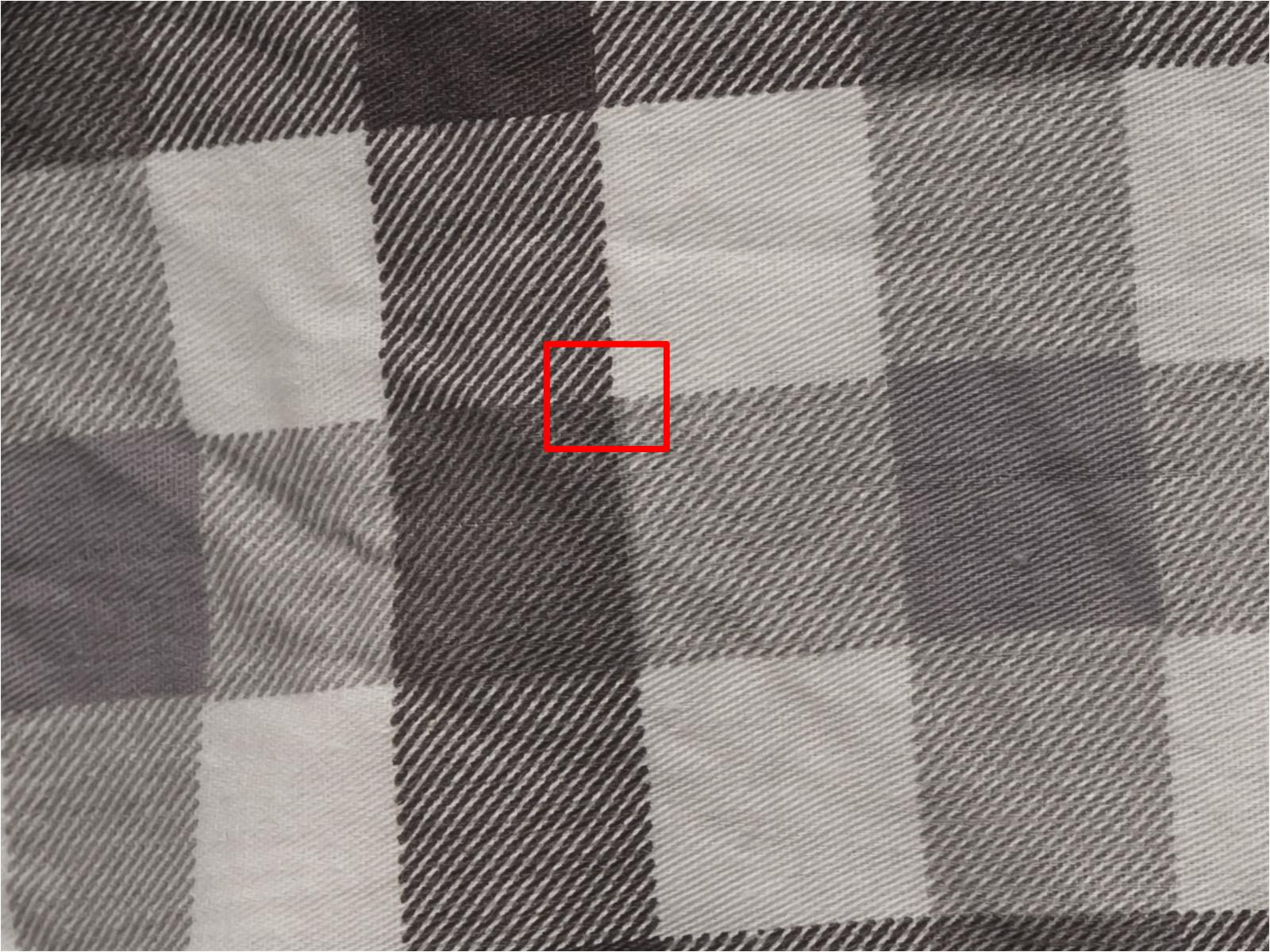}} \\
				towel
			\end{tabular}
		\end{adjustbox}
		\begin{adjustbox}{valign=t}
			\begin{tabular}{c@{\hspace{1mm}}c@{\hspace{1mm}}c@{\hspace{1mm}}c}
				\subfloat{\includegraphics[width=0.16\textwidth,height=0.12\textwidth]{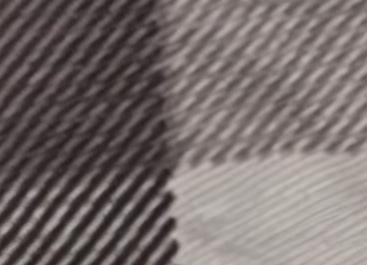}} 
				&
				\subfloat{\includegraphics[width=0.16\textwidth,height=0.12\textwidth]{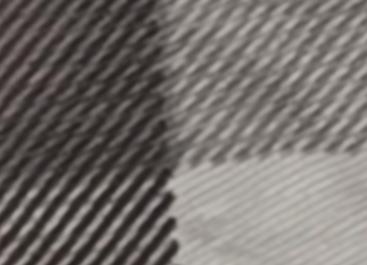}} 
				&
         		\subfloat{\includegraphics[width=0.16\textwidth,height=0.12\textwidth]{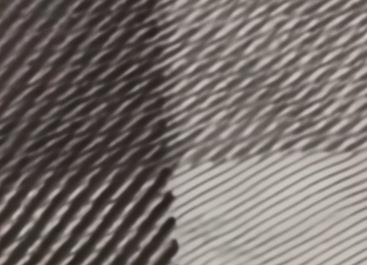}} 
				&
         		\subfloat{\includegraphics[width=0.16\textwidth,height=0.12\textwidth]{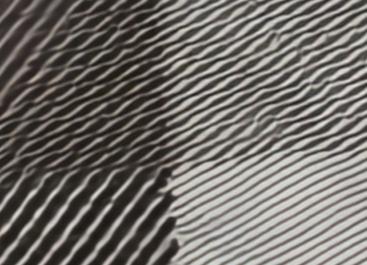}} 
			    \\
         		ESRGAN-Bic \cite{DBLP:conf/eccv/WangYWGLDQL18} & RCAN-Bic \cite{DBLP:conf/eccv/ZhangLLWZF18}  & RCAN-Real \cite{DBLP:conf/eccv/ZhangLLWZF18} & RealSR \cite{cai2019toward} \\ 
         		\specialrule{0em}{-8pt}{0pt}
         		\subfloat{\includegraphics[width=0.16\textwidth,height=0.12\textwidth]{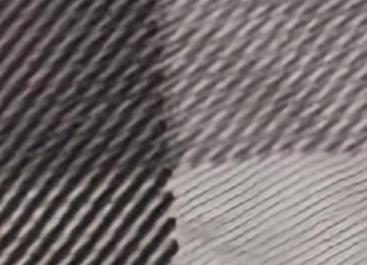}} 
         		&
         		\subfloat{\includegraphics[width=0.16\textwidth,height=0.12\textwidth]{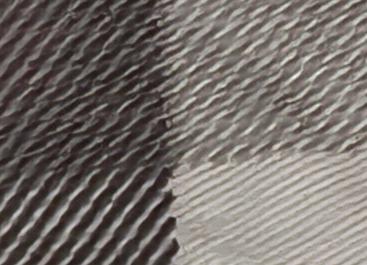}} 
         		&
         		\subfloat{\includegraphics[width=0.16\textwidth,height=0.12\textwidth]{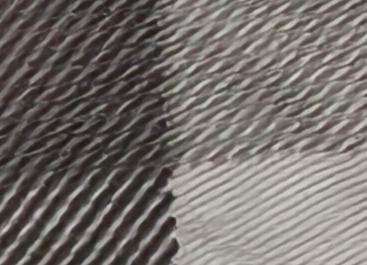}} 
         		&
         		\subfloat{\includegraphics[width=0.16\textwidth,height=0.12\textwidth]{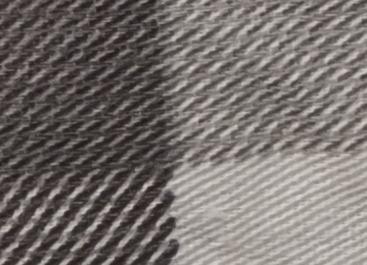}} \\
         		CinCGAN \cite{DBLP:conf/cvpr/YuanLZZDL18} & Ours-NoOver & Ours & Real HR Image \\  
			\end{tabular}
		\end{adjustbox}           
    \end{tabular}
    }
    
    \scalebox{0.8}{
    \begin{tabular}{cc}
    	\begin{adjustbox}{valign=t}
			\begin{tabular}{c}
			\specialrule{0em}{-8pt}{0pt}
				\subfloat{\includegraphics[width=0.23\textwidth,height=0.26\textwidth]{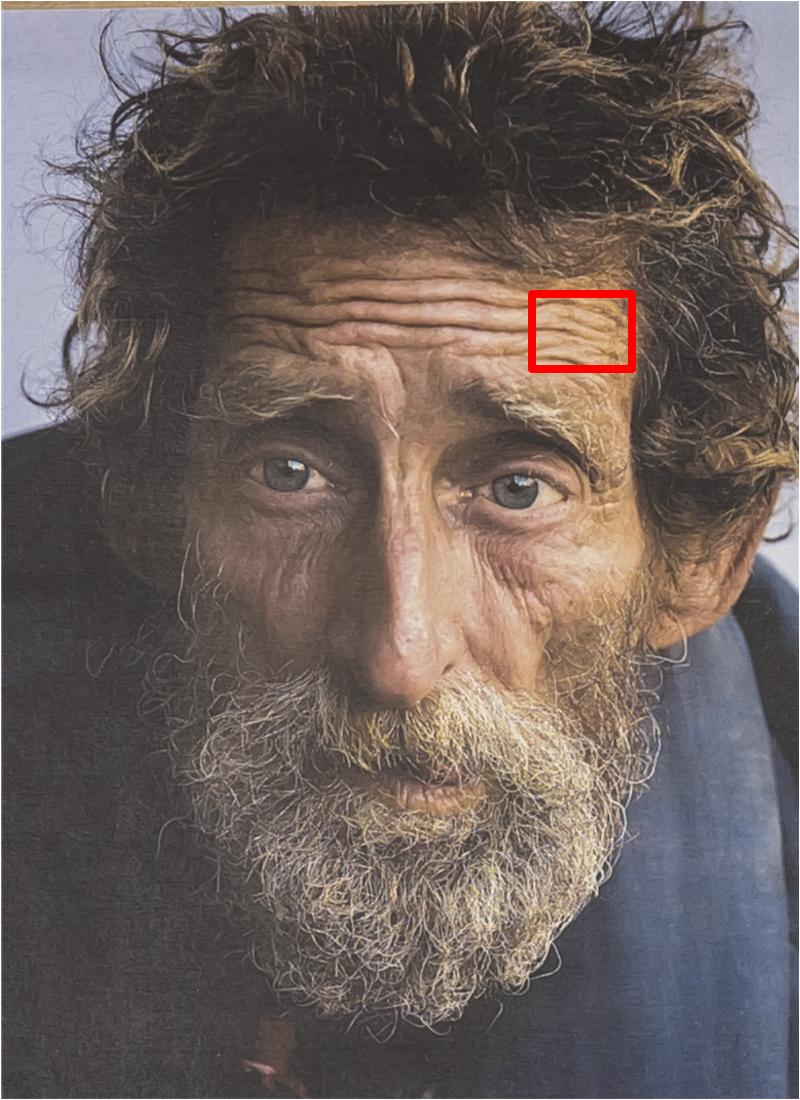}} \\
				man
			\end{tabular}
		\end{adjustbox}
		\begin{adjustbox}{valign=t}
			\begin{tabular}{c@{\hspace{1mm}}c@{\hspace{1mm}}c@{\hspace{1mm}}c}
			\specialrule{0em}{-8pt}{0pt}
				\subfloat{\includegraphics[width=0.16\textwidth,height=0.12\textwidth]{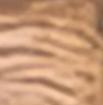}} 
				&
				\subfloat{\includegraphics[width=0.16\textwidth,height=0.12\textwidth]{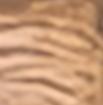}} 
				&
         		\subfloat{\includegraphics[width=0.16\textwidth,height=0.12\textwidth]{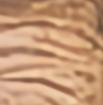}} 
         		&
         		\subfloat{\includegraphics[width=0.16\textwidth,height=0.12\textwidth]{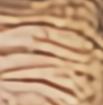}} 
				\\
         		ESRGAN-Bic \cite{DBLP:conf/eccv/WangYWGLDQL18} & RCAN-Bic \cite{DBLP:conf/eccv/ZhangLLWZF18}    & RCAN-Real \cite{DBLP:conf/eccv/ZhangLLWZF18} & RealSR \cite{cai2019toward}  \\
         		\specialrule{0em}{-8pt}{0pt}
         		\subfloat{\includegraphics[width=0.16\textwidth,height=0.12\textwidth]{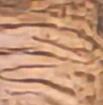}} 
         		&
         		\subfloat{\includegraphics[width=0.16\textwidth,height=0.12\textwidth]{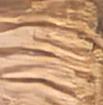}} 
         		&
         		\subfloat{\includegraphics[width=0.16\textwidth,height=0.12\textwidth]{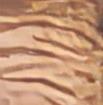}} 
         		&
         		\subfloat{\includegraphics[width=0.16\textwidth,height=0.12\textwidth]{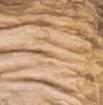}} \\
         		CinCGAN \cite{DBLP:conf/cvpr/YuanLZZDL18} & Ours-NoOver & Ours & Real HR Image \\  
			\end{tabular}
			
		\end{adjustbox}           
    \end{tabular}
    }
    \vspace{-0.3cm}
    \caption{Visual comparisons for 4x SR on the testing data from RealSR \cite{cai2019toward}. Local areas full of details and textures are zoom in. Our approach can produce better visual perceptual quality with more clean and natural textures.}
    \vspace{-0.5cm}
    \label{fig:visual_RealSR}
\end{figure*}
 
\par
We implement $\mathcal{G}$ and $\mathcal{F}$ as CNNs with eight residual blocks \cite{DBLP:conf/cvpr/LimSKNL17}, respectively. Global residual skip connection is applied between input and output, which helps to stabilize training and maintain color consistency between input and output. 
For the discriminators \dgnote{ $\cD_\tsyn$ and $\cD_\treal$}, we use the four-layer architecture in PatchGan \cite{DBLP:conf/cvpr/IsolaZZE17,DBLP:conf/eccv/LiW16}.

\subsection{Degradation Adaptive SR Network}
\label{sec:deg_adap_sr_net}
After obtaining the aligned image pair set $\{(I_\tgen^L, I_\treal^H)\}$, we train the real-world SR network $\cR(\cdot)$ under paired supervision. 
We adopt RCAN \cite{DBLP:conf/eccv/ZhangLLWZF18} as our baseline network. 
As shown in Figure \ref{fig:flowmap}, we use the pixel-wise content loss $\ell_1$-loss and the relative adversarial average GAN (RaGAN) loss \cite{DBLP:conf/eccv/WangYWGLDQL18} on the SR results $\cR(I_\tgen^L)$:

\begin{equation}
    \mathcal{L}_{\ell_1} (\cR) = \mathbb{E}_{I_\treal^H \sim  p(I_\treal^H)} 
    \left \| \cR(I_{\tgen}^{L}) - I_\treal^H  \right \|_{1}, 
\label{eq_L1}
\end{equation}
\begin{equation}
    \begin{aligned}
    \mathcal{L}_\text{RaGAN} (\cR, & \cD_\text{Ra}) = - \mathbb{E}_{I_{\tgen}^{L} \sim  p(I_{\tgen}^{L})} [\log (\mathcal{D}_\text{Ra} (\cR(I_{\tgen}^{L}), I_\treal^H) ) ] \\
    &- \mathbb{E}_{I_\treal^H \sim  p(I_\treal^H)} 
    [\log (1 - \mathcal{D}_\text{Ra} (I_\treal^H, \cR(I_{\tgen}^{L})) ) ],
    \end{aligned}
\label{eq_RaGan}
\end{equation}
where $p(I_\treal^H)$ and $p(I_{\tgen}^{L})$ denote the data distributions of $I_\treal^H$ and $I_{\tgen}^{L}$, $\mathcal{D}_\text{Ra}$ is the relativistic average discriminator \cite{DBLP:conf/eccv/WangYWGLDQL18, DBLP:conf/iclr/Jolicoeur-Martineau19}.

\begin{figure*}[!t]
    \scriptsize
	\centering
	\begin{tabular}{c@{\hspace{0.05mm}}c@{\hspace{0.05mm}}c@{\hspace{0.05mm}}c@{\hspace{0.05mm}}c@{\hspace{0.05mm}}c@{\hspace{0.05mm}}c@{\hspace{0.05mm}}c}
		\subfloat{\includegraphics[width=0.14\textwidth,height=0.12\textwidth]{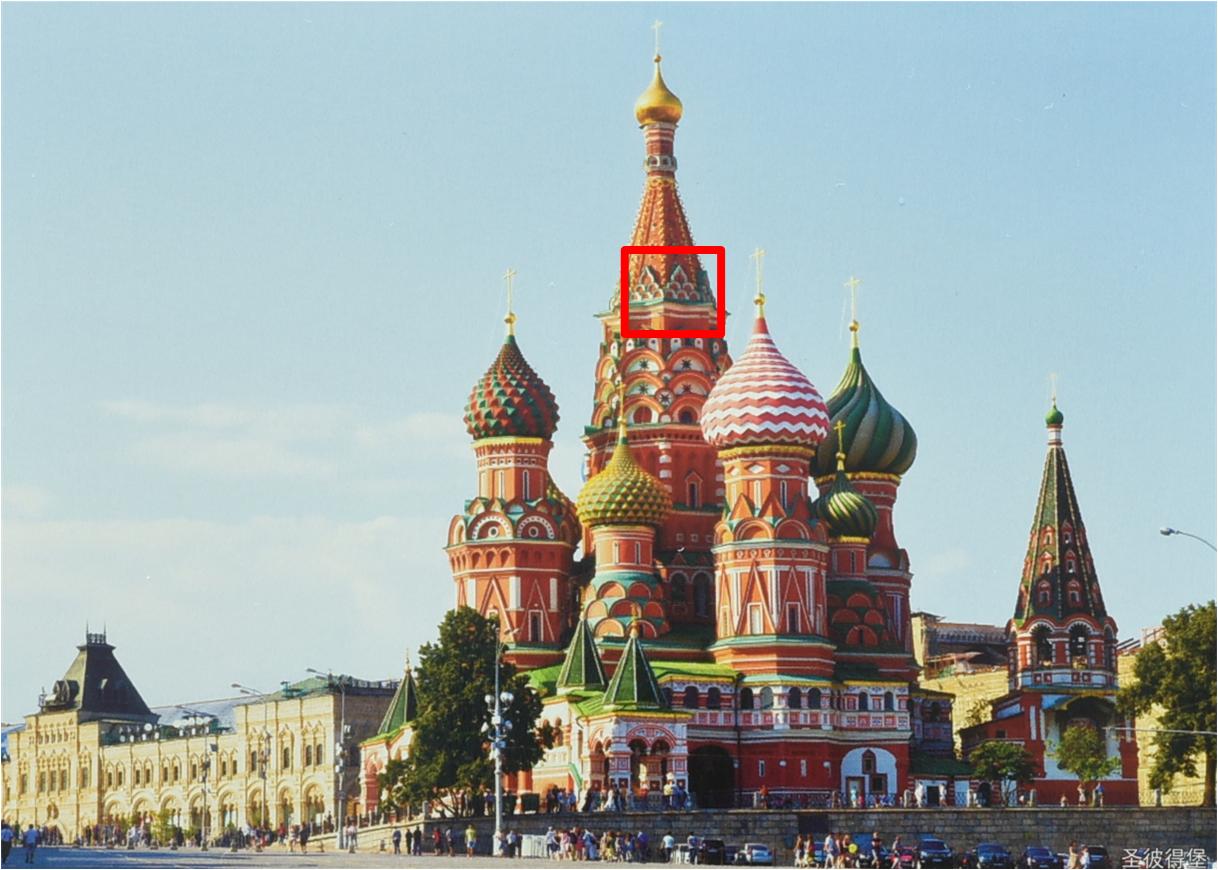}}
		&
		\subfloat{\includegraphics[width=0.12\textwidth,height=0.12\textwidth]{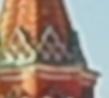}} 
		&
		\subfloat{\includegraphics[width=0.12\textwidth,height=0.12\textwidth]{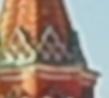}} 		
		&
        \subfloat{\includegraphics[width=0.12\textwidth,height=0.12\textwidth]{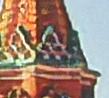}} 
        &
        \subfloat{\includegraphics[width=0.12\textwidth,height=0.12\textwidth]{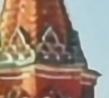}} 
        &
		\subfloat{\includegraphics[width=0.12\textwidth,height=0.12\textwidth]{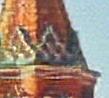}}
        &	
        \subfloat{\includegraphics[width=0.12\textwidth,height=0.12\textwidth]{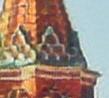}}           
        &
        \subfloat{\includegraphics[width=0.12\textwidth,height=0.12\textwidth]{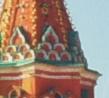}} \\
        
        \specialrule{0em}{-8pt}{0pt}
        \subfloat{\includegraphics[width=0.14\textwidth,height=0.12\textwidth]{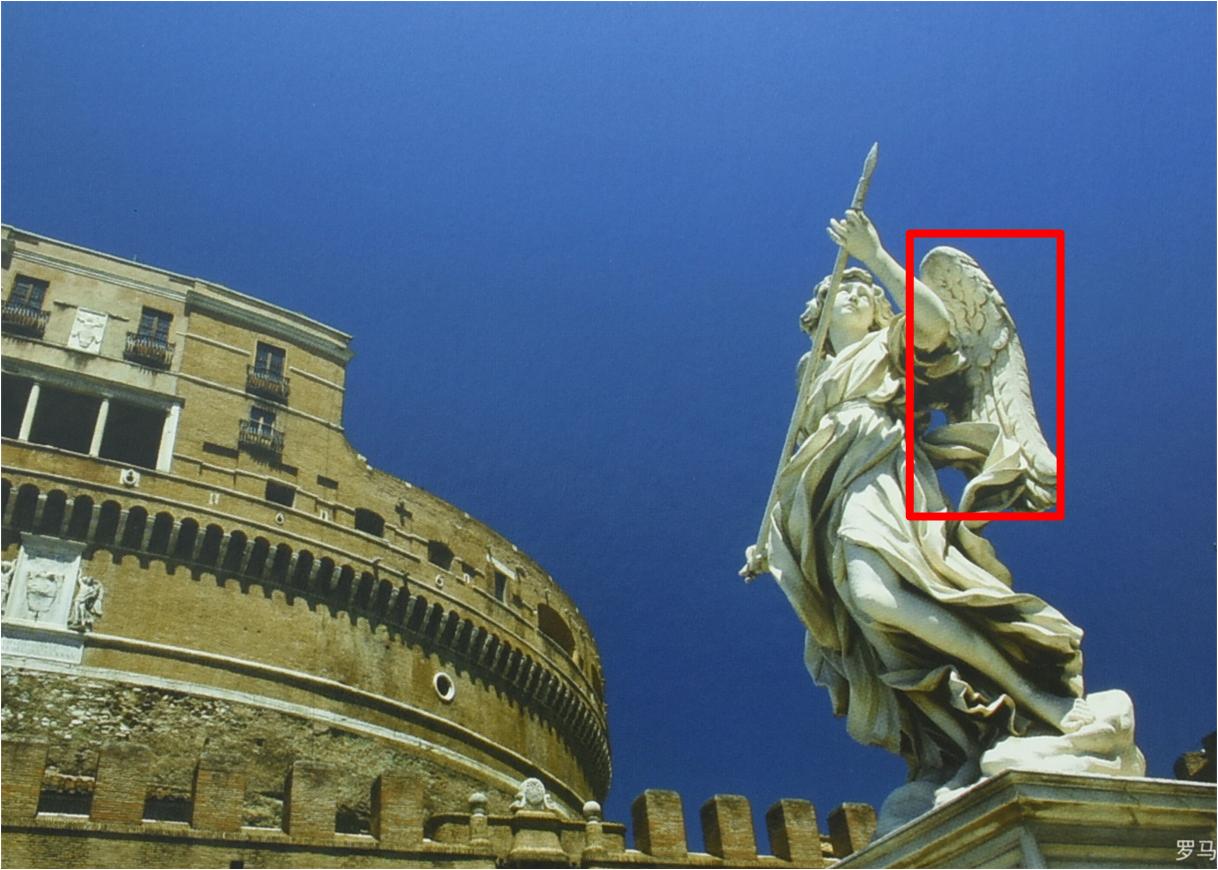}}
		&
		\subfloat{\includegraphics[width=0.12\textwidth,height=0.12\textwidth]{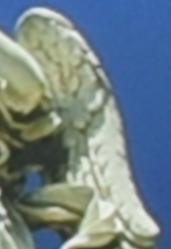}} 
		&
		\subfloat{\includegraphics[width=0.12\textwidth,height=0.12\textwidth]{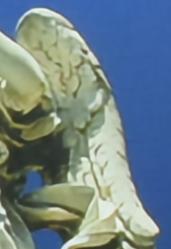}} 		
		&
        \subfloat{\includegraphics[width=0.12\textwidth,height=0.12\textwidth]{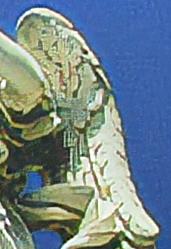}} 
        &
        \subfloat{\includegraphics[width=0.12\textwidth,height=0.12\textwidth]{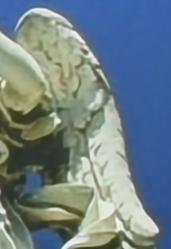}} 
        &
        \subfloat{\includegraphics[width=0.12\textwidth,height=0.12\textwidth]{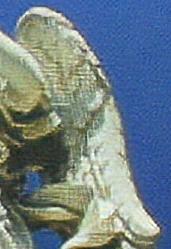}}
		&
        \subfloat{\includegraphics[width=0.12\textwidth,height=0.12\textwidth]{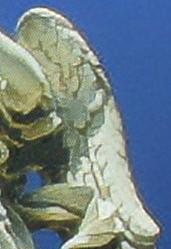}}           
        &
        \subfloat{\includegraphics[width=0.12\textwidth,height=0.12\textwidth]{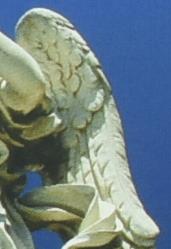}} \\
         		
		~& RCAN-Bic \cite{DBLP:conf/eccv/ZhangLLWZF18} & RCAN-Real \cite{DBLP:conf/eccv/ZhangLLWZF18} & CamSR-SRGAN \cite{DBLP:conf/cvpr/ChenXTZW19} & CamSR-VDSR \cite{DBLP:conf/cvpr/ChenXTZW19} & CinCGAN \cite{DBLP:conf/cvpr/YuanLZZDL18}  & Ours & Real HR Image
    \end{tabular}
    \vspace{-0.2cm}
    \caption{Visual comparisons for 3x SR on the testing data from City100 dataset \cite{DBLP:conf/cvpr/ChenXTZW19}. The proposed network gets rid of blurry and artifacts, it produces high-quality super-resolved images, especially edges and textures.}
    \vspace{-0.5cm}
    \label{fig:visual_city100}
\end{figure*}

\par
\noindent \textbf{Degradation adaptive loss functions.}
If we assume the domain gap between $\{I_\tgen^L\}$ and  $\{I_\treal^L\}$ has been removed as previous methods \cite{DBLP:conf/cvpr/YuanLZZDL18,DBLP:conf/eccv/BulatYT18,DBLP:journals/corr/abs-1909-09629}, we can stop here and apply the trained $\cR(\cdot)$ to handle real LR images. However, although the domain gap has been reduced, it is hard to achieve the ideal case. 
To enable the SR network trained with $\{I_\tgen^L\}$ can adapt to real-world LR degradation seen in testing better, we further alleviate the influence of domain while training SR network. 
As shown in Figure \ref{fig:flowmap}, we let $\cR(\cdot)$ take both $I^L_\tgen$ and $I^L_\treal$ as inputs (only) during training.
Since the paired ground truth for $I^L_\treal$ is unavailable, we thus use an adversarial loss between $\cR(I^L_\treal)$ and $I^H_\treal$:
\begin{equation}
    \begin{split}
    \mathcal{L}_\text{GAN-real}(\cR, & \mathcal{D}_\text{HR}) = \mathbb{E}_{I_{\treal}^{H} \sim  p(I_{\treal}^{H})} 
    [\log (\mathcal{D}_\text{HR} (I_{\treal}^{H}) ) ] \\
    & \mathbb{E}_{I_{\treal}^{L} \sim  p(I_{\treal}^{L})} 
    [\log (1-\mathcal{D}_\text{HR} (\cR(I_{\treal}^{L}) )) ],
    \end{split}
\label{eq_d1}
\end{equation}
where $\cD_\text{HR}$ denotes adversarial discriminator attempting to match the distribution of output image $\cR(I^L_\treal)$ and $I^H_\treal$.
We propose an adaptive loss to align the network responses of $I^L_\tgen$ and $I^L_\treal$, which is designed as a adversarial loss between the features of $I^L_\tgen$ and $I^L_\treal$: 
\begin{equation}
    \begin{aligned}
    \mathcal{L}_\text{ada}(\widehat{\cR},\mathcal{D}_\text{ada}) &= \mathbb{E}_{I_{\tgen}^{L} \sim  p(I_{\tgen}^{L})} 
    [\log (\mathcal{D}_\text{ada} (\widehat{\cR}(I_{\tgen}^{L})) ) ] \\
    & \mathbb{E}_{I_{\treal}^{L} \sim  p(I_{\treal}^{L})} 
    [\log (1-\mathcal{D}_\text{ada} (\widehat{\cR}(I_{\treal}^{L}) )) ],
    \end{aligned}
\label{eq_d2}
\end{equation}
where $\widehat{\cR}(\cdot)$ represents the network intermediate feature extraction and $\cD_\text{ada}$ denotes \dgnote{the domain discriminator of each low-level feature region produced by $\widehat{\cR}(\cdot)$}. The effectiveness of similar loss function has also been proven in domain adaptation learning tasks \cite{shen2019regularizing}. The two discriminators are implemented with the structure of PatchGAN as above \cite{DBLP:conf/cvpr/IsolaZZE17,DBLP:conf/eccv/LiW16}. 

\par
The full objective for degradation adaptive SR network is:
\begin{equation}
\begin{aligned}
    \mathcal{L}_\text{SR} & = \lambda_{1}  \mathcal{L}_{\ell_1} + \lambda_{2} \mathcal{L}_\text{RaGAN} \\
    & +\lambda_{3}  \mathcal{L}_\text{GAN-real}(\cR,\mathcal{D}_\text{HR}) + \lambda_{4} \mathcal{L}_\text{ada}(\widehat{\cR},\mathcal{D}_\text{ada}),
\end{aligned}
\label{eq_totalloss_SR}
\end{equation}
\dgnote{where $\lambda_{1}$, $\lambda_{2}$, $\lambda_{3}$, $\lambda_{4}$ are weigthts of each loss.}

\par
During traing, the whole framework is free from any form of the direct supervision from $I^L_\treal$ and $I^H_\treal$.

\section{Experiments}

\subsection{Experimental Settings}
\noindent \textbf{Training and testing data.}
The proposed model is trained on unpaired real-world LR and HR image sets and tested on the corresponding real-world LR images. 
We conduct experiments relying on three real-world SR datasets, RealSR \cite{cai2019toward}, City100 \cite{DBLP:conf/cvpr/ChenXTZW19} and SR-RGB \cite{DBLP:conf/cvpr/ZhangCNK19}. 
RealSR contains 500 LR-HR image pairs ($80\%$ for training) taken by Canon and Nikon cameras for $\times 4$ SR. City100 contains 100 real LR and HR images, respectively, taken with a Nikon camera for $\times 3$ SR, where 95 images are for training. SR-RGB contains 498 non-aligned LR and HR images, respectively, taken by Sony cameras for $\times 4$ SR, where 448 images are set as training data. 
We train our model on their training set and evaluate it on the testing set. 
Although LR-HR pairs are provided in these datasets (and are pre-registered in RealSR and City100), we ignore the pair information in our experiments. 
In the stricter ``non-overlapping'' setting, we randomly drop part of LR and HR training images to ensure no overlapping contents between LR and HR set.

\noindent \textbf{Evaluation Metrics.} To quantitatively evaluate different methods, we use PSNR and Structural Similarity index (SSIM) \cite{DBLP:journals/tip/WangBSS04}. PSNR and SSIM are based solely on distortion measurement and fail to account for many nuances of human perception \cite{DBLP:conf/cvpr/ZhangIESW18,DBLP:conf/cvpr/BlauM18,DBLP:conf/eccv/BlauMTMZ18}. Therefore, the widely used LPIPS \cite{DBLP:conf/cvpr/ZhangIESW18} and perception index (PI) \cite{DBLP:conf/eccv/BlauMTMZ18} are also chose for quantitative perceptual evaluation. LPIPS is a learned metric for the perceptual similarity between recovered and ground-truth (GT) images. PI is a no-reference perceptual quality measurement, and it is highly correlated with the subjective evaluation of human observers \cite{DBLP:conf/eccv/BlauMTMZ18}.

\subsection{Training Details}
The degradation generation network $\cG(\cdot)$ and the SR Network $\cR(\cdot)$ are trained separately. For training $\cG(\cdot)$, the parameters in Eq.(\ref{eq_totalloss_gan}) are set to be $w_{1}=2$, $w_{2}=2$ and $w_{3}=0.5$, respectively. 
For training the SR network, we set the parameters in Eq.(\ref{eq_totalloss_SR}) as $\lambda_{1}=1$, $\lambda_{2}=0.1$ (for RealSR and City100, and $\lambda_{2}=0.001$ for SR-RGB), $\lambda_{3}=1$ and $\lambda_{4}=2$.
For training both $\cG(\cdot)$ and $\cR(\cdot)$, we use Adam optimizer and set the batch size and learning rate as 8 and $1 \times 10^{-4}$ for all the models, respectively. Given training images, LR RGB patches with a size of $64 \times 64$ are extracted as inputs. The learning rate decreases to half every 1600K iterations.

\begin{figure*}[!t]
    \scriptsize
	\centering
	\begin{tabular}{c@{\hspace{0.05mm}}c@{\hspace{0.05mm}}c@{\hspace{0.05mm}}c@{\hspace{0.05mm}}c@{\hspace{0.05mm}}c@{\hspace{0.05mm}}c@{\hspace{0.05mm}}c}
		\subfloat{\includegraphics[width=0.14\textwidth,height=0.12\textwidth]{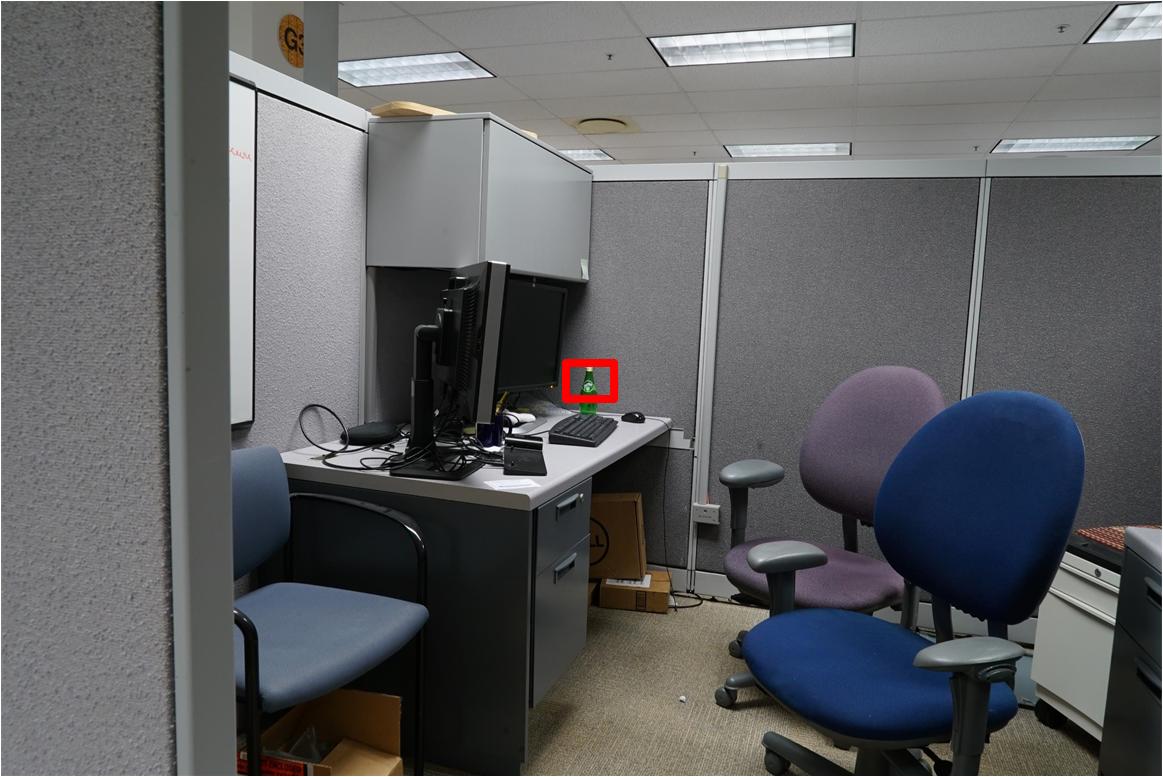}}
		&
		\subfloat{\includegraphics[width=0.12\textwidth,height=0.12\textwidth]{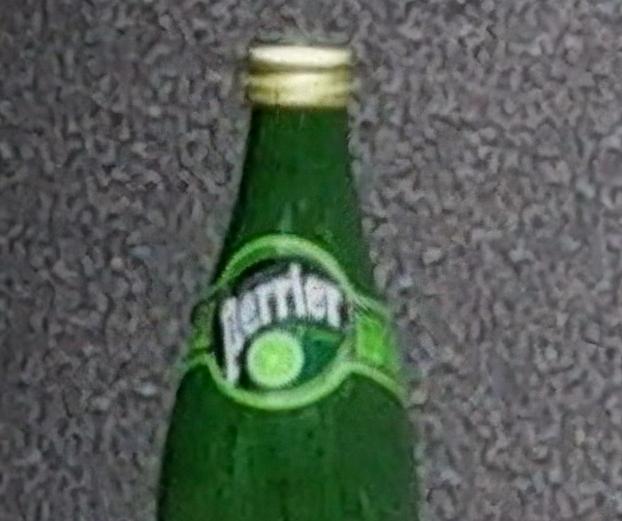}} 
		&
		\subfloat{\includegraphics[width=0.12\textwidth,height=0.12\textwidth]{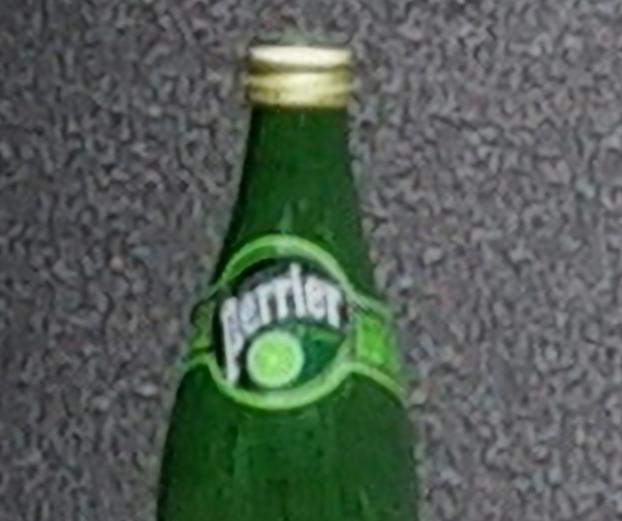}} 		
		&
        \subfloat{\includegraphics[width=0.12\textwidth,height=0.12\textwidth]{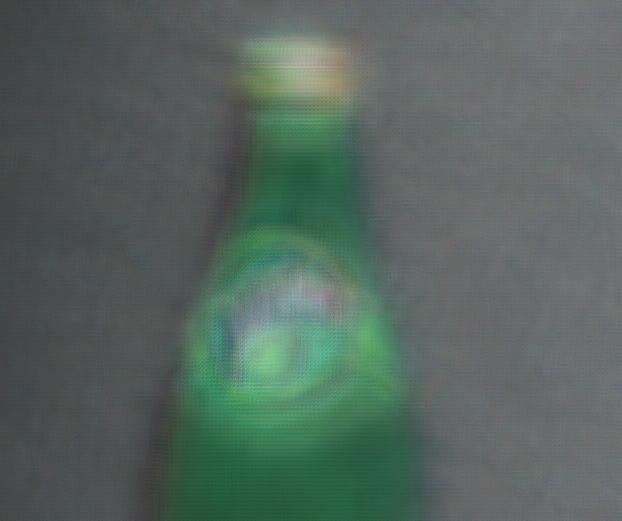}} 
        &
        \subfloat{\includegraphics[width=0.12\textwidth,height=0.12\textwidth]{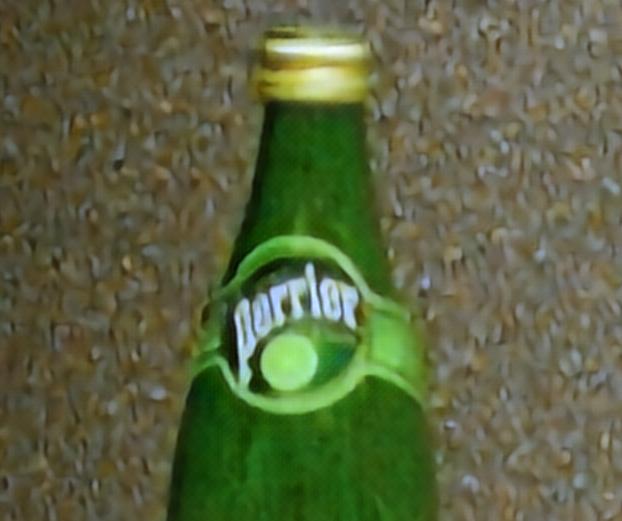}} 
        &
		\subfloat{\includegraphics[width=0.12\textwidth,height=0.12\textwidth]{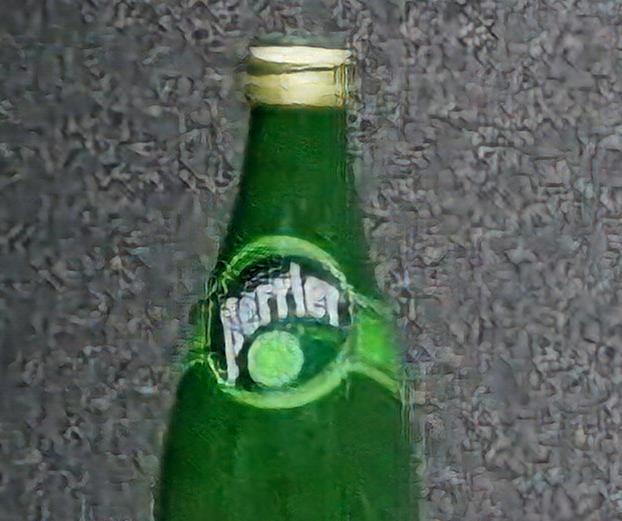}}
        &	
        \subfloat{\includegraphics[width=0.12\textwidth,height=0.12\textwidth]{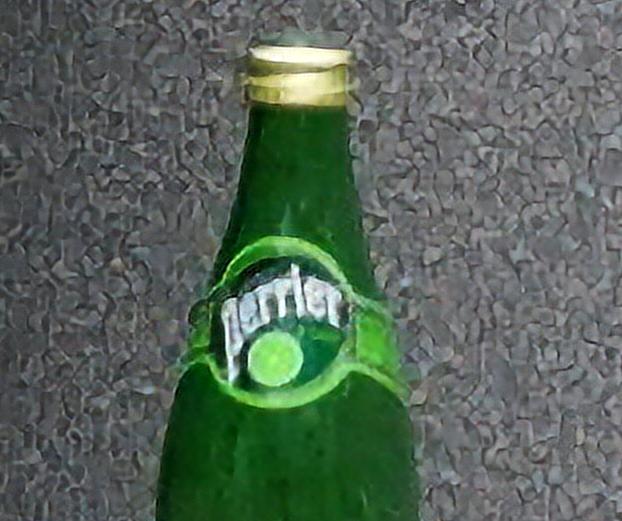}}           
        &
        \subfloat{\includegraphics[width=0.12\textwidth,height=0.12\textwidth]{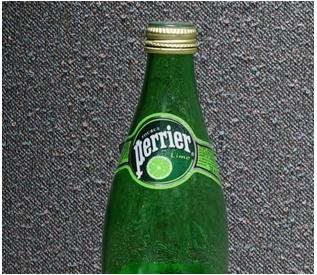}} \\
        
        \specialrule{0em}{-8pt}{0pt}
        \subfloat{\includegraphics[width=0.14\textwidth,height=0.12\textwidth]{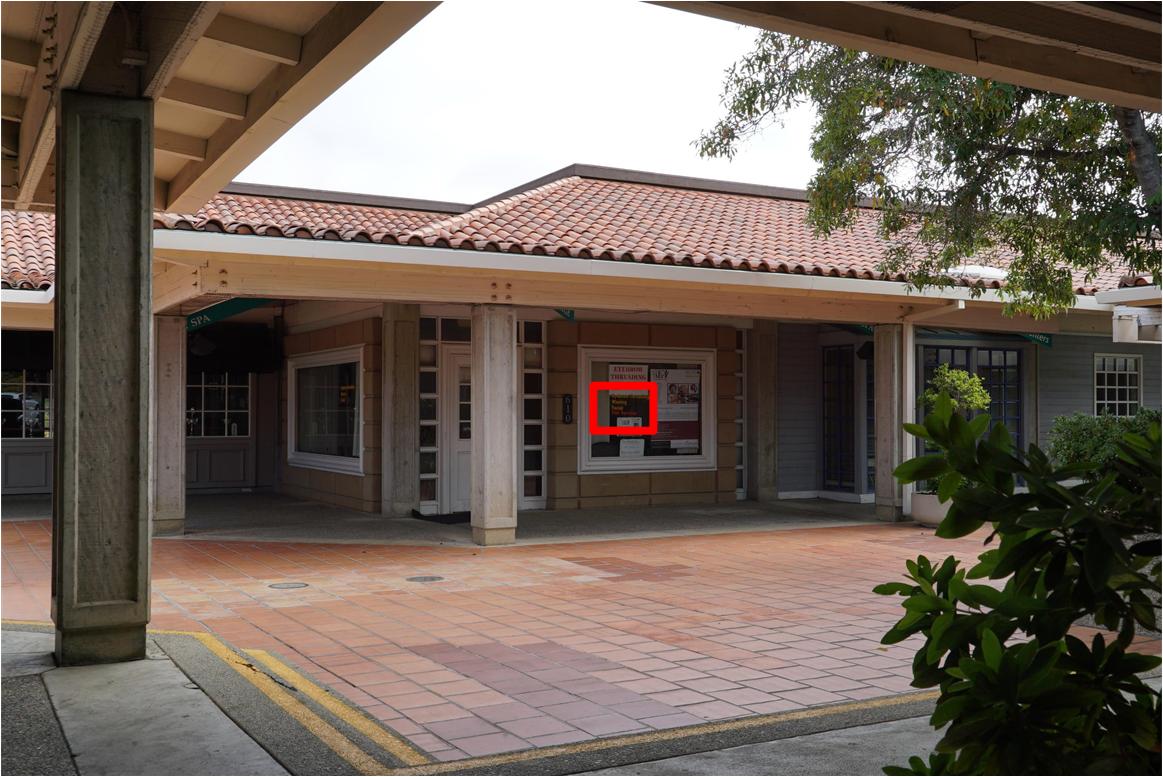}}
		&
		\subfloat{\includegraphics[width=0.12\textwidth,height=0.12\textwidth]{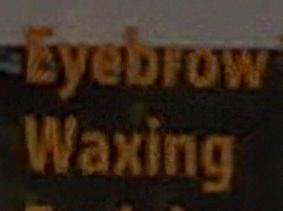}} 
		&
		\subfloat{\includegraphics[width=0.12\textwidth,height=0.12\textwidth]{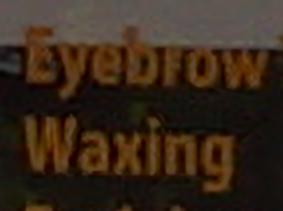}} 		
		&
        \subfloat{\includegraphics[width=0.12\textwidth,height=0.12\textwidth]{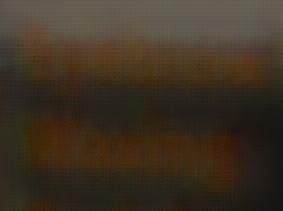}} 
        &
        \subfloat{\includegraphics[width=0.12\textwidth,height=0.12\textwidth]{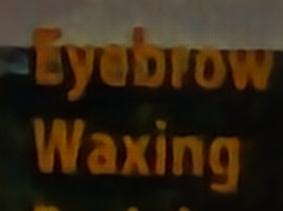}} 
        &
        \subfloat{\includegraphics[width=0.12\textwidth,height=0.12\textwidth]{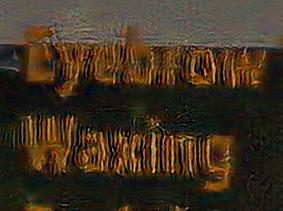}}
		&
        \subfloat{\includegraphics[width=0.12\textwidth,height=0.12\textwidth]{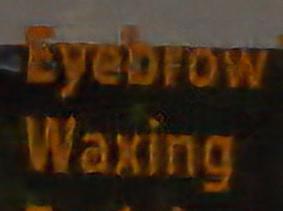}}           
        &
        \subfloat{\includegraphics[width=0.12\textwidth,height=0.12\textwidth]{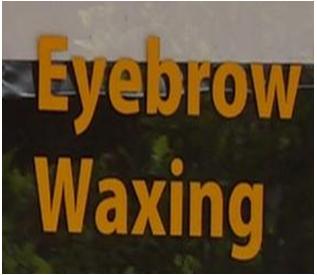}} \\
         		
		~& ESRGAN-Bic \cite{DBLP:conf/eccv/WangYWGLDQL18} & RCAN-Bic \cite{DBLP:conf/eccv/ZhangLLWZF18} & RCAN-Real \cite{DBLP:conf/eccv/ZhangLLWZF18} & ZoomSR \cite{DBLP:conf/cvpr/ZhangCNK19} & CinCGAN \cite{DBLP:conf/cvpr/YuanLZZDL18}   & Ours & Real HR Image
    \end{tabular}
    \vspace{-0.2cm}
    \caption{Visual comparisons for 4x SR on the testing data from SR-RGB dataset \cite{DBLP:conf/cvpr/ZhangCNK19}. Misalignment exists in LR-HR pairs as no pre-registration is applied. The non-aligned real HR image is shown for reference. 
    Our approach recovers more informative information and finer details.}
    \label{fig:visual_SR-RAW}
    \vspace{-0.3cm}
\end{figure*}

\begin{table}[!t]
\begin{center}
\begin{adjustbox}{max width=0.47\textwidth}
\begin{tabular}{c|c|c|c||c|c}
\hline
~ & Methods & $\uparrow$ PSNR        & $\uparrow$ SSIM  & $\downarrow$ LPIPS & $\downarrow$ PI    \\ \hline

\multirow{4}{*}{\rotatebox{90}{paired}} & ESRGAN-Bic \cite{DBLP:conf/eccv/WangYWGLDQL18}              & 27.569      & 0.774     & 0.415 & 7.442     \\

~& RCAN-Bic \cite{DBLP:conf/eccv/ZhangLLWZF18}   & 27.647      & 0.780     & 0.442   & 7.844   \\ \cline{2-6} 

~& RCAN-Real  \cite{DBLP:conf/eccv/ZhangLLWZF18}               & 29.486      & 0.831     & 0.276   & 7.179   \\ 

~& RealSR \cite{cai2019toward}                & 28.702      & 0.821     & 0.276   & 7.019   \\ \hline

\multirow{3}{*}{\rotatebox{90}{unpa.}} & CinCGAN \cite{DBLP:conf/cvpr/YuanLZZDL18} & 27.447 & 0.762 & 0.290 & 5.705  \\ \cline{2-6} 

~& Ours-NoOver            & 27.606      & 0.766     & 0.257  & 5.343   \\  
~& Ours                    & 27.723      & 0.786     & 0.251  &  5.242  \\ \hline
\end{tabular}
\end{adjustbox}
\end{center}
\vspace{-0.5cm}
\caption{Quantitative comparisons for 4x SR on RealSR dataset \cite{cai2019toward}. ``paired'' and ``unpa.'' denote paired and unpaired methods, respectively. Our approach achieves the best results in terms of LPIPS and PI and comparable PSNR and SSIM values.  Note that a lower LPIPS and PI score indicates better image perceptual quality.}
\label{tab:RealSR}
\vspace{-0.5cm}
\end{table}

\subsection{Comparison with the State-of-the-art Methods} 
To validate the proposed method, we compare with the state-of-the-art SR methods based on both paired data and unpaired data, respectively. 
Specifically, we compare with the state-of-the-art deep learning SR methods ESRGAN \cite{DBLP:conf/eccv/WangYWGLDQL18} and our baseline model RCAN \cite{DBLP:conf/eccv/ZhangLLWZF18}, which are proposed to train and test on the bicubic degradation (paired) data; real-world SR methods RealSR \cite{cai2019toward}, CamSR-SRGAN \cite{DBLP:conf/cvpr/ChenXTZW19}, CamSR-VDSR \cite{DBLP:conf/cvpr/ChenXTZW19} and ZoomSR \cite{DBLP:conf/cvpr/ZhangCNK19}, which are proposed to handle the paired real-world SR data; and unpaired SR methods CinCGAN \cite{DBLP:conf/cvpr/YuanLZZDL18}. Although the code of \cite{DBLP:journals/corr/abs-1909-09629} is unavailable, it is similar to one of our ablation variant as discussed in Section \ref{sec:ablation}.

Based on ESRGAN and RCAN, we test how the models trained on synthetic (unrealistic) degradation work on real LR images. We generate synthetic SR datasets by applying bicubic degradation on HR images in the realistic datasets, and train the networks to obtain ESRGAN-Bic and RCAN-Bic. 
On each dataset, we train RCAN (denoted as RCAN-real) and the models proposed with the dataset based on the pre-registered LR-HR pairs. 
For above paired learning methods, the LR-HR pair information is directly used as supervision. 
For the unpaired methods, \ie, CinCGAN and our method, we ignore the pair relationship. 
CinCGAN is trained on the three real-world datasets using the same setup as in the paper.

\begin{table}[!t]
\begin{center}
\begin{adjustbox}{max width=0.47\textwidth}
\begin{tabular}{c|c|c|c||c|c}
\hline
~ & Methods &  $\uparrow$ PSNR        & $\uparrow$ SSIM      & $\downarrow$ LPIPS  & $\downarrow$ PI    \\ \hline 

\multirow{4}{*}{\rotatebox{90}{paired}} & RCAN-Bic  \cite{DBLP:conf/eccv/ZhangLLWZF18}               & 28.114      & 0.811      & 0.384  & 6.849   \\ \cline{2-6}

~ & RCAN-Real  \cite{DBLP:conf/eccv/ZhangLLWZF18}               & 30.016      & 0.864      & 0.260  & 5.881   \\ 
~& CamSR-SRGAN \cite{DBLP:conf/cvpr/ChenXTZW19}    & 25.257      & 0.764      & 0.195  & 3.450   \\ 
~& CamSR-VDSR  \cite{DBLP:conf/cvpr/ChenXTZW19}        & 30.260      & 0.868      & 0.263   & 5.825 \\ \hline

\multirow{2}{*}{\rotatebox{90}{unpa.}}& CinCGAN \cite{DBLP:conf/cvpr/YuanLZZDL18} & 26.221 & 0.703 & 0.303 & 3.783 \\ \cline{2-6}
~& Ours   & 27.790      & 0.789      & 0.184   & 3.697  \\ \hline
\end{tabular}
\end{adjustbox}
\end{center}
\vspace{-0.5cm}
\caption{Quantitative comparisons for 3x SR on City100 dataset \cite{DBLP:conf/cvpr/ChenXTZW19}. The unpaired method CinCGAN have a large gap with ours on all the evaluation metrics.}
\vspace{-0.5cm}
\label{tab:City100}
\end{table}

\noindent \textbf{Experiments on RealSR \cite{cai2019toward}.} 
Table \ref{tab:RealSR} shows quantitative comparisons among different approaches. 
The methods trained with bicubic degradation, \ie, ESRGAN-Bic and RCAN-Bic, do not perform well on realistic LR images, reflecting the influence of the domain gap. 
Relying on the pre-registration, the methods directly trained on the paired images, \ie, RCAN-Real and RealSR model, can obtain higher PSNR and SSIM values than the unpaired learning methods. 
The proposed unpaired learning methods achieve the best performances under the perceptual evaluation metrics LPIPS and PI. 
The results are understandable.
The proposed unpaired method is trained using less supervisions, rendering better generalization for higher perceptual quality.
The other unpaired method CinCGAN is trained by only handling the domain gap in LR image domain. The performance is not better than ours.

\par
We show visual comparisons in Figure \ref{fig:visual_RealSR}, in which local areas full of details and textures are zoom in. The proposed method produces results containing clean and natural textures, which are more similar to the real HR image. The results of the paired learning methods RealSR and RCAN-Real contain blurred patterns and fack textures, maybe because the misalignment influences training.

\begin{table}[!t]
\begin{center}
\begin{adjustbox}{max width=0.47\textwidth}
\begin{tabular}{c|c|c|c||c|c}
\hline
~& Methods & $\uparrow$ PSNR  & $\uparrow$ SSIM  & $\downarrow$ LPIPS  & $\downarrow$ PI     \\ \hline
\multirow{4}{*}{\rotatebox{90}{paired}} & ESRGAN-Bic \cite{DBLP:conf/eccv/WangYWGLDQL18}   & 16.703      & 0.542     & 0.534  & 5.385   \\ 
~& RCAN-Bic \cite{DBLP:conf/eccv/ZhangLLWZF18}  & 16.713      & 0.554     & 0.583   & 7.210   \\ \cline{2-6}
~& RCAN-Real \cite{DBLP:conf/eccv/ZhangLLWZF18} & 17.887      & 0.601     & 0.779   & 8.569   \\ 
~& ZoomSR  \cite{DBLP:conf/cvpr/ZhangCNK19} & 17.086      & 0.570     & 0.519    & 6.532  \\ \hline
\multirow{3}{*}{\rotatebox{90}{unpa.}} & CinCGAN \cite{DBLP:conf/cvpr/YuanLZZDL18} & 16.136 & 0.498 & 0.538 & 4.826 \\ \cline{2-6}
~& Ours-CroCam     & 16.008            & 0.460          &  0.520    & 4.852  \\
~& Ours     & 16.234            & 0.542          &  0.491    & 4.647    \\ \hline
\end{tabular}
\end{adjustbox}
\end{center}
\vspace{-0.5cm}
\caption{Quantitative comparisons for 4x SR on SR-RGB \cite{DBLP:conf/cvpr/ZhangCNK19}.
No pre-registration is applied. Misalignment exists in LR-HR pairs. 
}
\vspace{-0.3cm}
\label{tab:SR-RAW}
\end{table}

\noindent \textbf{Stricter ``non-overlapping'' setting.}
In the used datasets \cite{cai2019toward}, paired LR-HR images are provided $\{(I^L_{\treal, i}, I^H_{\treal,i})\}_{i=1}^N$. We conduct unpaired training by directly separate the LR and HR image sets as $\mbL=\{I^L_{\treal, i}\}_{i=1}^N$ and $\mbH=\{I^H_{\treal, i}\}_{i=1}^N$, which is the basic setting in our experiments. 
Considering that $\mbL$ and $\mbH$ contains shared contents in LR and HR images, to further validate the generalization of the proposed method, we build a stricter setting with non-overlapping contents between training sets for LR and HR. 
Specifically, we drop part of images (with different indexes) in $\mbL$ and $\mbH$ to get $\mbH_\text{no}=\{I^H_{\treal, i}\}_{i=1}^{N'}$ and $\mbL_\text{no}=\{I^L_{\treal, i}\}_{i=N'+1}^{N}$ for training. 

Based on RealSR dataset with $N=400$, we obtain non-overlapping training sets by letting $N'=300$ and train our model on it.
We show the results in Table \ref{tab:RealSR} (denoted as Ours-NoOver). We can see our model still can obtain good results using only half samples of the basic setting used by other methods. Our results under the basic setting and non-overlapping setting are similar, which shows the generalization of our method but does not down weigh the value of the basic setting.

\begin{table*}[!t]
\begin{center}
\begin{adjustbox}{max width=0.6\textwidth}
\begin{tabular}{c|c|c|c|c|c|c|c|c}
\hline
\multirow{2}{*}{Methods} & \multicolumn{4}{c|}{Degradation} & \multicolumn{4}{c}{Super-Resolution} \\ \cline{2-9} 
                        & $\uparrow$ PSNR       & $\uparrow$ SSIM     & $\downarrow$ LPIPS  & $\downarrow$ PI   & $\uparrow$ PSNR        & $\uparrow$ SSIM       & $\downarrow$ LPIPS   & $\downarrow$ PI  \\
                        \hline
RCAN-Bic \cite{DBLP:conf/eccv/ZhangLLWZF18}   & -     & -    & -  & -   & 27.647      & 0.780     & 0.442   & 7.844  \\ \hline
Ours-Bic   & -     & -    & -  & -   & 26.830      & 0.718     & 0.297 & 5.283  \\ 
Ours-GAN           & 16.053     & 0.804    & 0.219  & 5.262  & 19.615      & 0.740      & 0.310  & 5.342   \\ 
Ours-OneCycle          & 34.786     & 0.963    & 0.051 & 5.168   & 27.485      & 0.770      & 0.280    & 5.265  \\ 
Ours                    & 35.508     & 0.965    & 0.047  & 5.214  & 27.723      & 0.786      & 0.251    & 5.242  \\ \hline
\end{tabular}
\end{adjustbox}
\end{center}
\vspace{-0.5cm}
\caption{Ablation study on the degradation network based on RealSR dataset \cite{cai2019toward}.}
\label{tab:Ablation1}
\vspace{-0.6cm}
\end{table*}

\noindent \textbf{Experiments on City100  \cite{DBLP:conf/cvpr/ChenXTZW19}.} 
City100 is captured and pre-processed in the ways similar to RealSR. We show the results in Table \ref{tab:City100}. The methods proposed in the paper, \ie, CamSR-SRGAN and CamSR-VDSR, are trained under the image pair supervision. Supervised RCAN-Real and CamSR-VDSR obtain higher PSNR and SSIM values but lower scores on perceptual evaluation. CamSR-SRGAN improves the perceptual quality but obtains lower scores on PSNR and SSIM than others. The unpaired method CinCGAN have a large gap with ours on the quantitative comparison. In Figure \ref{fig:visual_city100}, over-enhanced effects appear on the estimated results of CamSR-SRGAN. Ours approach gets rid of blurry and artifacts, it produces sharper edges and clear textures.

\noindent \textbf{Experiments on SR-RGB \cite{DBLP:conf/cvpr/ZhangCNK19}.} 
SR-RGB dataset is the sRGB version of the SR-RAW dataset in \cite{DBLP:conf/cvpr/ZhangCNK19}, in which the LR-HR image pairs are not specifically pre-registered. The results are shown in Table \ref{tab:SR-RAW}. 

\par
\noindent \textbf{Local distortion metrics vs. perceptual quality.}
Before analyzing the results, we would like to discuss the observed contradictions between the local distortion metrics (\eg, PSNR and SSIM) and the perceptual quality measurements (\eg, LPIPS, PI and visual quality). 
In this experiment, we directly train the pixel-wise-loss-based method RCAN-Real on the nonaligned SR-RGB dataset. It surprisingly obtains very high PSNR and SSIM values. But we observe that the output HR images are extremely blurry, due to the non-aligned training pairs. 
This reflects the mismatch between the PSNR and the real visual quality \cite{DBLP:conf/cvpr/ZhangIESW18,DBLP:conf/eccv/BlauMTMZ18}. 
Similar results have also been observed in the experiments on RealSR and City100 as discussed above. 
These observations imply that the perceptual measurements \ie, LPIPS and PI, can reflect the real image qualities, which are consistent with the visualizations.

\par
The ZoomSR method proposed in \cite{DBLP:conf/cvpr/ZhangCNK19} applies a specifically designed loss to handle the misalignment, which obtains good quantitative and visual results.
As shown in both Table \ref{tab:SR-RAW} and Figure \ref{fig:visual_SR-RAW}, the proposed method achieves the best perceptual quality, which recovers more informative information and finer details. 

\par 
\noindent \textbf{Generalization cross camera.} We apply our model trained on RealSR to test on the real LR images in SR-RGB. It produces results on-par with the model trained on SR-RGB (denoted as Ours-CroCam), which shows the proposed method has the ability to generalize across cameras.

\par
More experimental results are left in Appendix \ref{sec:app_more_res}.

\subsection{Ablation Studies}
\label{sec:ablation}
We conduct ablation studies on RealSR dataset.

\noindent \textbf{Study on the degradation network.}
We study the importance of each part in the degradation generation network by testing different model variants and show the results in Table \ref{tab:Ablation1}. 
We first show the importance of our whole degradation network by training our SR network only using the LR images generated by bicubic degradation, referred to as Ours-Bic. It also shows that the proposed SR network is more powerful than the based RCAN. The ablation studies show that the proposed degradation network can mimic the real LR images well and produce high-quality HR results under the unpaired learning setting.
We also study the model variants of only using GAN loss (\ie, Ours-GAN) and using single cycle structure (\ie, Ours-OneCycle) and show the quality of the generated LR images and the SR results in Table \ref{tab:Ablation1}.

\begin{table}[htp]
\begin{center}
\begin{adjustbox}{max width=0.40\textwidth}
\begin{tabular}{c|c|c||c|c}
\hline
 Methods&  $\uparrow$ PSNR        & $\uparrow$ SSIM      & $\downarrow$ LPIPS  & $\downarrow$ PI  \\ \hline 
RCAN-Bic \cite{DBLP:conf/eccv/ZhangLLWZF18}                & 27.647      & 0.780     & 0.442 &  7.844  \\ \hline
w/ only $\ell_1$-loss            & 28.383      & 0.799     & 0.281 & 6.572    \\ 
w/o RaGAN           & 27.543      & 0.754     & 0.277 & 5.673   \\ 
w/o ada. loss       & 27.468      & 0.745     & 0.275 & 6.039    \\
Ours                   & 27.723      & 0.786     & 0.251 & 5.242    \\ \hline
\end{tabular}
\end{adjustbox}
\end{center}
\vspace{-0.5cm}
\caption{Ablation study on the loss functions in our SR network based on RealSR dataset \cite{cai2019toward}.}
\label{tab:Ablation2}
\vspace{-0.4cm}
\end{table}

\noindent \textbf{Study on loss functions of SR network.} 
We study the behaviours of the loss function used in our SR network. 
The results of the SR networks trained with only $\ell_1$-loss, without RaGAN, and without the proposed adaptive degradation loss are shown in Table \ref{tab:Ablation2}, respectively. Note that ``w/o ada. loss'' is similar to the method in \cite{DBLP:journals/corr/abs-1909-09629}. Compared with RCAN-Bic, when adding our degradation generation network, it successfully improves the  performance.
The results show that the adversarial loss functions (including $\cL_\text{RaGAN}$, $\cL_\text{GAN-real}(\cR,\cD_\text{HR})$ and $\cL_\text{ada}(\widehat{\cR},\cD_\text{ada})$) are important for high perceptual quality. 
Specifically, the proposed degradation adaptive loss is more important than RaGAN for handling domain shift and recovering high-quality visual details. 

\section{Conclusion}
In this paper, we propose a method to learn real-world SR from a set of unpaired LR and HR images. Given a set of unpaired LR-HR images, we firstly train a degradation generation network to generate realistic LR images by capturing their distribution. We then further minimize the discrepancy between the generated data and real data while learning a degradation adaptive SR network. We show that unpaired learning methods can do comparably or better than paired learning methods in a realistic setting. The proposed unpaired method achieves state-of-the-art SR results on real-world images, even in the datasets that favor the paired-learning methods more.

\section{Appendix}
\subsection{Visualization of Different Diagrams}
In this section, we summarize the main idea of different methods by visualizing the main diagrams and show the differences among them. 

\par
\begin{figure*}[]
 \footnotesize
	\centering
	\begin{tabular}{c}

		\subfloat{\includegraphics[width=0.61\textwidth]{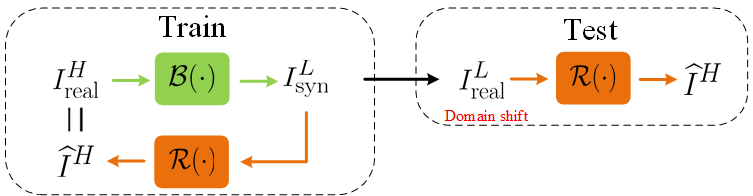}}\\
		(a) Learning SR based on synthetic bicubic LR degradation \cite{DBLP:conf/eccv/WangYWGLDQL18,DBLP:conf/eccv/ZhangLLWZF18} 
		\\
		\specialrule{0em}{-8pt}{0pt}
		\subfloat{\includegraphics[width=0.7\textwidth]{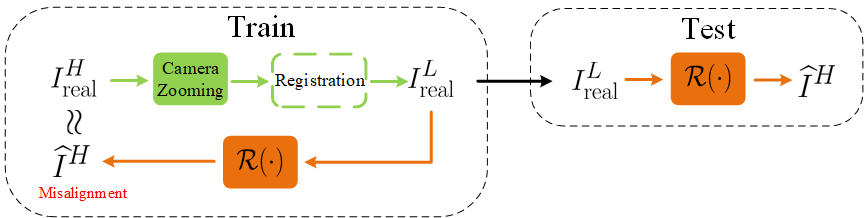}} \\
         (b) Learning SR based on paired images taken by adjusting focal length of the cameras \cite{cai2019toward,DBLP:conf/cvpr/ChenXTZW19,DBLP:conf/cvpr/ZhangCNK19}\\ \hline
        \specialrule{0em}{-8pt}{0pt}
        \subfloat{\includegraphics[width=0.61\textwidth]{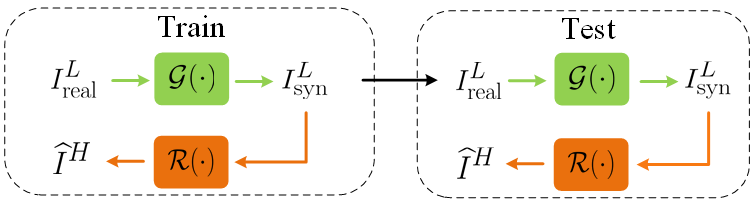}}
		\\
		(c) Unpaired Learning by generating synthetic LR images from realistic LR images \cite{DBLP:conf/cvpr/YuanLZZDL18}\\
		\specialrule{0em}{-8pt}{0pt}
		\subfloat{\includegraphics[width=0.77\textwidth]{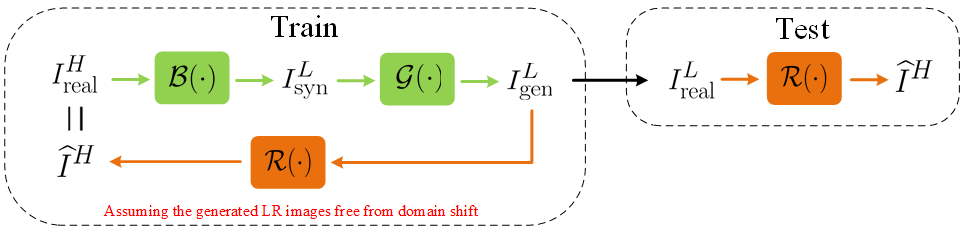}} \\ 
        (d) Unpaired learning by generating realistic LR images from HR images \cite{DBLP:journals/corr/abs-1909-09629}\\
        \specialrule{0em}{-8pt}{0pt}
		\subfloat{\includegraphics[width=0.77\textwidth]{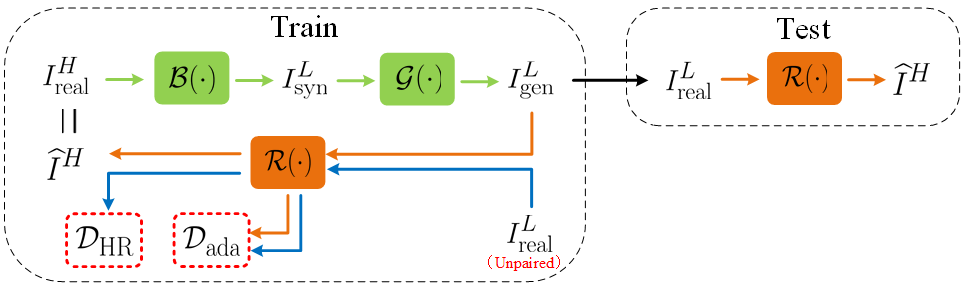}} \\ 
        (e) The proposed method: unpaired learning by generating realistic LR images (from HR images) and adapting degradation in SR.

    \end{tabular}
    \caption{Diagrams of the main ideas of five different paradigms applying for real-world SR.  (a) and (b) show the paired learning-based SR methods, based on synthetic bicubic LR degradation and captured real-world LR-HR image pairs. 
    (c), (d) and (e) show the unpaired learning-based methods. Before learning to do SR, they first learn to obtain paired supervision signals for SR in different ways. Different methods are used to alleviate the influence of the domain gap. In the figures, we illustrate how the SR network $\mathcal{R}(\cdot)$ is trained (left) and applied during inference. The aligned supervisions are indicated by ``$=$''. 
    ``$\approx $'' is used to indicate the misaligned LR-HR pairs. 
    $\mathcal{B}(\cdot)$ and $\mathcal{G}(\cdot)$ denote the bicubic degradation and the trainable generation network, respectively.}
    \label{fig:basic_methods}
\end{figure*}

Figure \ref{fig:basic_methods} illustrates the core ideas of the proposed method and four other main approaches for real-world SR, including paired learning \cite{DBLP:conf/eccv/WangYWGLDQL18,DBLP:conf/eccv/ZhangLLWZF18,cai2019toward,DBLP:conf/cvpr/ChenXTZW19,DBLP:conf/cvpr/ZhangCNK19} and unpaired learning methods \cite{DBLP:conf/cvpr/YuanLZZDL18,DBLP:journals/corr/abs-1909-09629}. 
In Figure \ref{fig:basic_methods} (a), we show the bicubic LR degradation based baseline model \cite{DBLP:conf/eccv/WangYWGLDQL18,DBLP:conf/eccv/ZhangLLWZF18}. The model trained with synthetic bicubic LR images cannot well generalize to the realistic data due to the domain gap. 
Figure \ref{fig:basic_methods} (b) shows the other kind of paired learning-based works \cite{cai2019toward,DBLP:conf/cvpr/ChenXTZW19,DBLP:conf/cvpr/ZhangCNK19}. 
To avoid the influence of the domain gap, LR-HR image pairs are captured by adjusting the focal length for learning SR network. Even after pre-registration, the LR-HR pairs still suffer from misalignment, which is unavoidable and influences the learning of SR network.

\par
Figure \ref{fig:basic_methods} (c), (d) and (e) show the paradigms of different unpaired learning. 
The approach shown in Figure \ref{fig:basic_methods} (c) \cite{DBLP:conf/cvpr/YuanLZZDL18} tends to handle real-world LR image by learning to map the real LR images to synthetic bicubic LR images, which can be seen as a pre-processing operation for SR. 
It is learned by training to transfer the real LR images to bicubic images. 
The unpaired bicubic degradation LR images are used to guide the process of generation. Then SR network $\mathcal{R(\cdot)}$ directly uses the artificially synthetic LR-HR image pairs for training as in Figure \ref{fig:basic_methods} (a). 
Figure \ref{fig:basic_methods} (d) shows another way to obtain paired supervision signals for SR network \cite{DBLP:journals/corr/abs-1909-09629}. The approach learns to generate realistic images $I_\text{gen}^{L}$ by mimicking the real LR degradation with the guidance of unpaired realistic LR images. 
The method for obtaining real LR images (for maintaining paired LR and HR images for training) is similar to the proposed method. 
By assuming that the generated LR images are free from domain shift with realistic LR images, it then utilizes the perfectly aligned LR-HR training dataset $\{ (I_\text{gen}^{L},I_\text{real}^{H}) \}$ to supervised train the SR network $\mathcal{R}(\cdot)$. 
Figure \ref{fig:basic_methods} (e) shows the main idea of the proposed method. 
As shown in Figure \ref{fig:basic_methods} (e), instead of assuming the generated LR images $I_\text{gen}^{L}$ free from domain shift as in (d), we propose to further align the model response on the generated and realistic LR images with degradation adaptive losses while training our SR network. We let $\mathcal{R}(\cdot)$ take both $I_\text{gen}^{L}$ and unpaired $I_\text{real}^{L}$ as inputs (only) during training. The details have been described in Section \ref{sec:deg_adap_sr_net}.

\subsection{More Experimental Results}
\label{sec:app_more_res}
\begin{figure*}[!t]
\footnotesize
	\centering
	\begin{tabular}{c@{\hspace{0.5mm}}c@{\hspace{0.5mm}}c}

		\subfloat{\includegraphics[trim=50 50 50 50, clip,width=0.22\textwidth,height=0.15\textwidth]{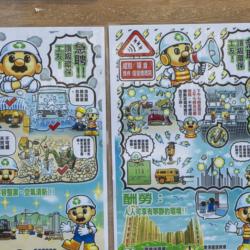}}
		&
		\subfloat{\includegraphics[trim=50 50 50 50, clip,width=0.22\textwidth,height=0.15\textwidth]{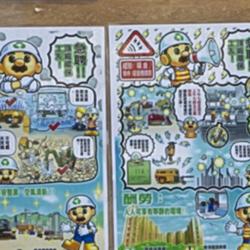}}
		&
		\subfloat{\includegraphics[trim=50 50 50 50, clip,
		width=0.22\textwidth,height=0.15\textwidth]{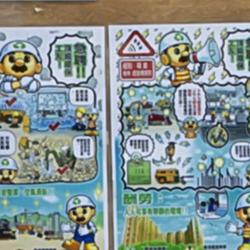}}\\
		
		\specialrule{0em}{-10pt}{0pt}
		\subfloat{\includegraphics[trim=50 0 50 100, clip,width=0.22\textwidth,height=0.15\textwidth]{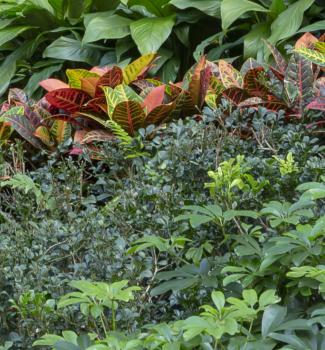}}
		&
		\subfloat{\includegraphics[trim=50 0 50 100, clip,width=0.22\textwidth,height=0.15\textwidth]{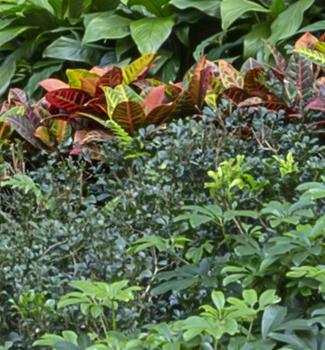}}
		&
		\subfloat{\includegraphics[trim=50 0 50 100, clip,width=0.22\textwidth,height=0.15\textwidth]{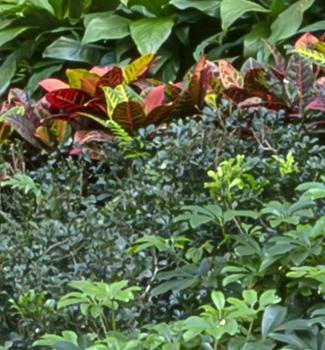}}\\
		
		 Bicubic Degradation LR Image & Our Generated LR Image & Real LR Image \\
		
		\multicolumn{3}{c}{(a) Examples from RealSR dataset \cite{cai2019toward}} \\
		
		\specialrule{0em}{-8pt}{0pt}
		\subfloat{\includegraphics[trim=100 100 100 100, clip,width=0.22\textwidth,height=0.15\textwidth]{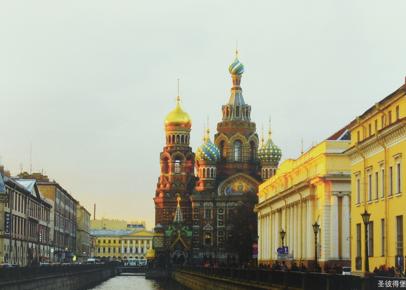}}
		&
		\subfloat{\includegraphics[trim=100 100 100 100, clip,width=0.22\textwidth,height=0.15\textwidth]{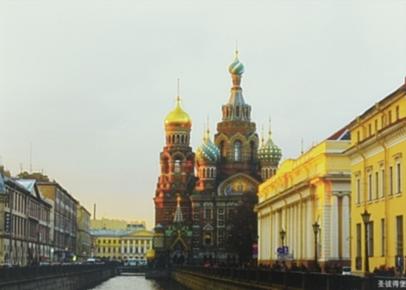}}
		&
		\subfloat{\includegraphics[trim=100 100 100 100, clip,width=0.22\textwidth,height=0.15\textwidth]{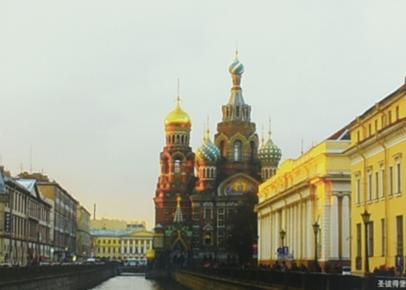}}\\
		\specialrule{0em}{-10pt}{0pt}
		
		\subfloat{\includegraphics[trim=150 120 130 50, clip,width=0.22\textwidth,height=0.15\textwidth]{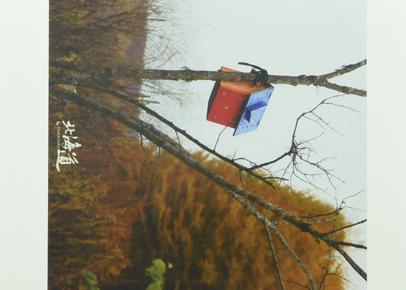}}
		&
		\subfloat{\includegraphics[trim=150 120 130 50, clip,width=0.22\textwidth,height=0.15\textwidth]{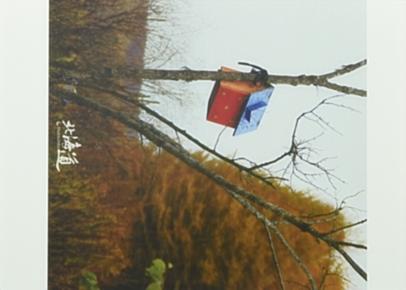}}
		&
		\subfloat{\includegraphics[trim=150 120 130 50, clip,width=0.22\textwidth,height=0.15\textwidth]{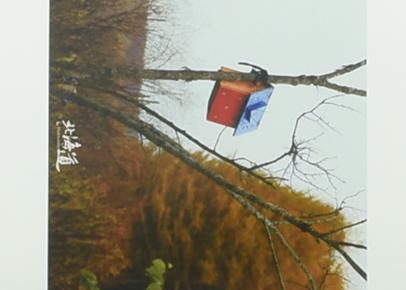}}\\
		
		 Bicubic Degradation LR Image & Our Generated LR Image & Real LR Image \\
		\multicolumn{3}{c}{(b) Examples from City100 dataset \cite{DBLP:conf/cvpr/ChenXTZW19}} \\
		
		\specialrule{0em}{-8pt}{0pt}
		
		\subfloat{\includegraphics[trim=100 100 100 100, clip,width=0.22\textwidth,height=0.15\textwidth]{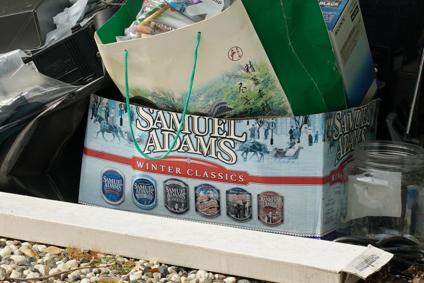}}
		&
		\subfloat{\includegraphics[trim=100 100 100 100, clip,width=0.22\textwidth,height=0.15\textwidth]{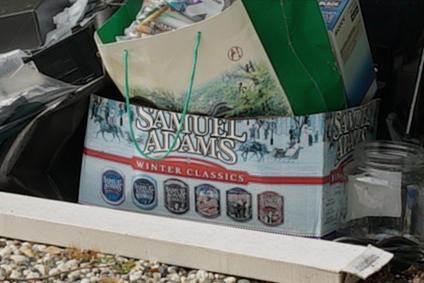}}
		&\subfloat{\includegraphics[trim=100 100 100 100, clip,width=0.22\textwidth,height=0.15\textwidth]{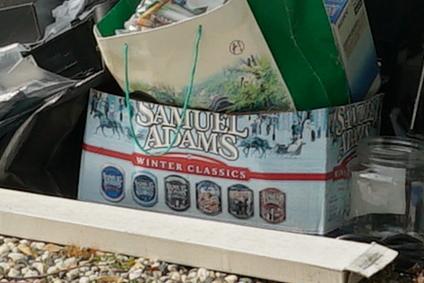}}
		\\
		\specialrule{0em}{-10pt}{0pt}
		\subfloat{\includegraphics[trim=100 50 100 150, clip,width=0.22\textwidth,height=0.15\textwidth]{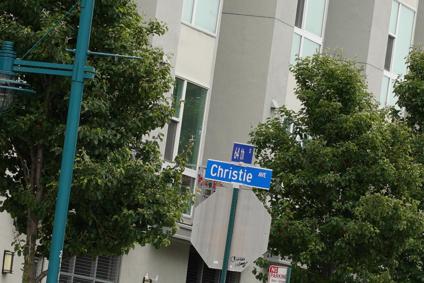}}
		&
		\subfloat{\includegraphics[trim=100 50 100 150, clip,width=0.22\textwidth,height=0.15\textwidth]{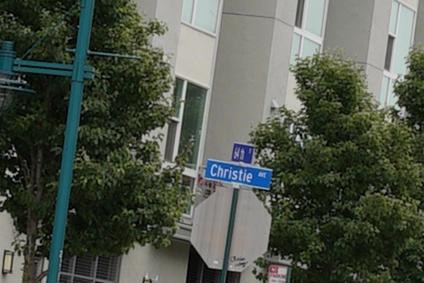}}
		&
		\subfloat{\includegraphics[trim=100 50 100 150, clip,width=0.22\textwidth,height=0.15\textwidth]{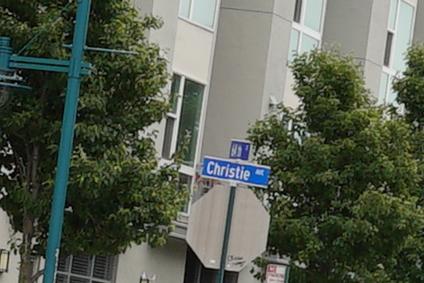}}
		\\
		Bicubic Degradation LR Image & Our Generated LR Image & Real LR Image \\
        \multicolumn{3}{c}{(c) Examples from SR-RGB dataset \cite{DBLP:conf/cvpr/ZhangCNK19}}
    \end{tabular}
    \vspace{-0.2cm}
    \caption{Visual results of our degradation generation network $\mathcal{G}(\cdot)$ on datasets \cite{cai2019toward,DBLP:conf/cvpr/ChenXTZW19,DBLP:conf/cvpr/ZhangCNK19}. 
    The bicubic degradation LR images, LR images generated by our method and the real LR images are illustrated. The LR images generated by our method can reflect characteristics of the realistic LR images. 
    }
    \vspace{-0.5cm}
    \label{fig:visual_degradation}
\end{figure*}

\noindent \textbf{Visual Results of the Degradation Generation Network.}
As described in Section 3.3 of the main paper, we learn a degradation generation network (\ie, $\cG(\cdot)$) that generates LR images from HR images by mimicking the real LR degradation.  
We show one example of the generated LR image in Figure 3 in the main paper. 
In this section, we show more results of the degradation generation network $\cG(\cdot)$ in Figure \ref{fig:visual_degradation}. 
In the realistic SR datasets \cite{cai2019toward,DBLP:conf/cvpr/ChenXTZW19,DBLP:conf/cvpr/ZhangCNK19}, (pre-registered or non-aligned) real LR-HR pairs are provided. We thus can use the corresponding real LR images as references to evaluate our generated LR images. Note that the pair information is used only in evaluation. 
Since there is still misalignment between the pairs, the differences to the real LR images are just for reference. 
As shown in Figure \ref{fig:visual_degradation}, the real-world LR images contain more complicated degradation than the LR images synthesized via bicubic degradation. 
Since bicubic LR images usually retain more details, the model trained on them cannot generalize well to real-world LR images in testing. 
Since the proposed degradation generation network is trained to fit the distribution of the real LR images, the LR images generated by our method reflect more characteristics of the realistic LR images.

\noindent \textbf{More Visual Comparisons with the State-of-the-art Methods.}
We show more visual results and the comparisons with previous state-of-the-art methods in Figure \ref{fig:visual_RealSR} and \ref{fig:visual_SR-RGB} in the following.
More examples from datasets RealSR \cite{cai2019toward} and SR-RGB \cite{DBLP:conf/cvpr/ZhangCNK19} are shown, which illustrate that the proposed method is general and stable to handle different cases.


\begin{figure*}[]
\footnotesize
	\centering
    
    \begin{tabular}{cc}
    	\begin{adjustbox}{valign=t}
			\begin{tabular}{c}
			\subfloat{\includegraphics[width=0.23\textwidth,height=0.17\textwidth]{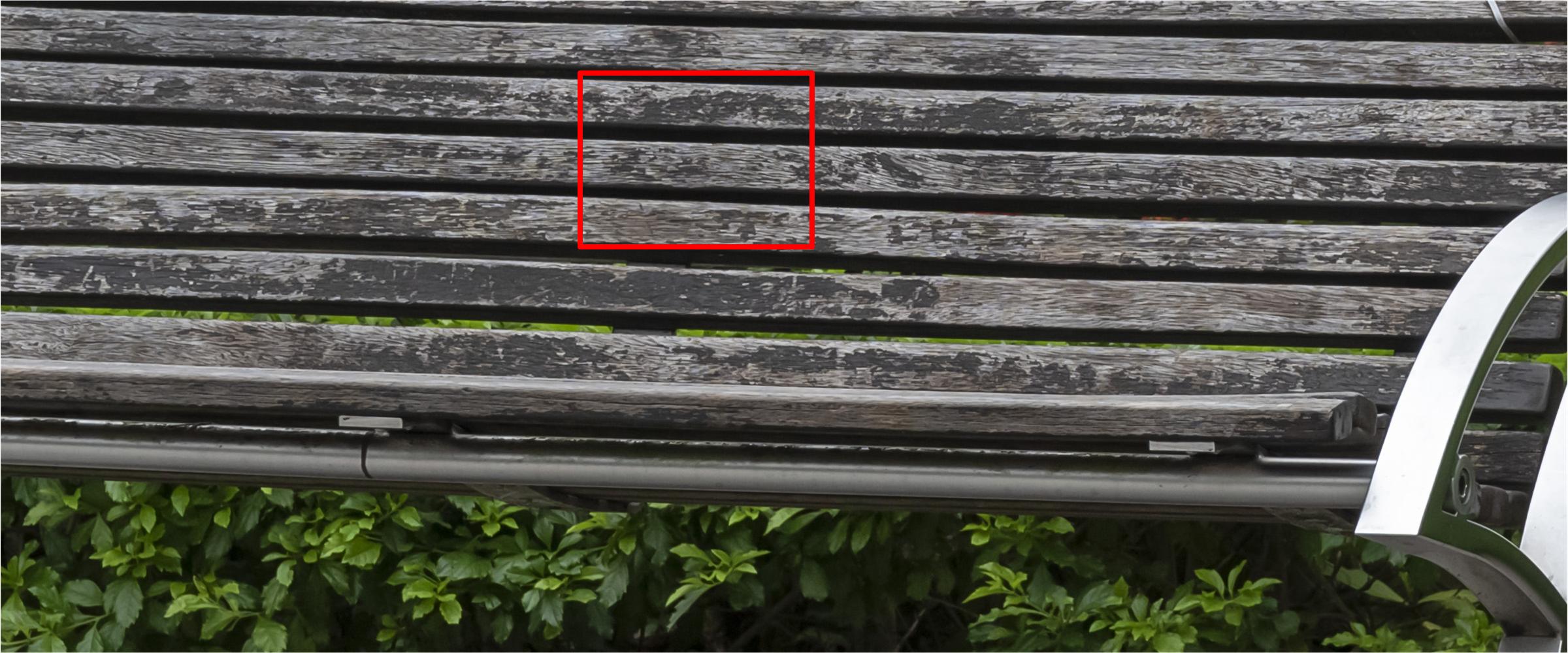}} \\
				bench
			\end{tabular}
		\end{adjustbox}
		\begin{adjustbox}{valign=t}
			\begin{tabular}{c@{\hspace{1mm}}c@{\hspace{1mm}}c@{\hspace{1mm}}c}
				\subfloat{\includegraphics[width=0.16\textwidth,height=0.12\textwidth]{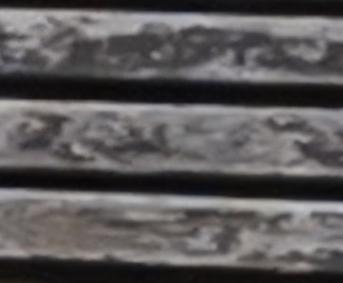}} 
				&
				\subfloat{\includegraphics[width=0.16\textwidth,height=0.12\textwidth]{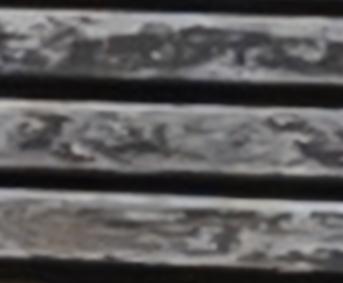}} 
				&
         		\subfloat{\includegraphics[width=0.16\textwidth,height=0.12\textwidth]{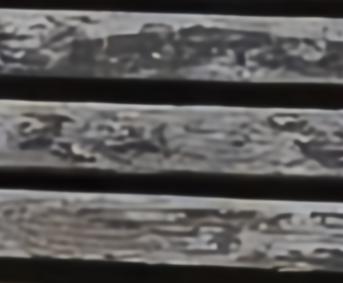}} 
         		&
				\subfloat{\includegraphics[width=0.16\textwidth,height=0.12\textwidth]{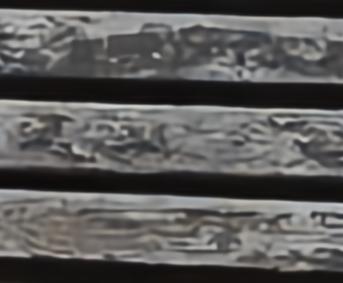}} 
         		\\
         		ESRGAN-Bic \cite{DBLP:conf/eccv/WangYWGLDQL18} & RCAN-Bic \cite{DBLP:conf/eccv/ZhangLLWZF18}  & RCAN-Real \cite{DBLP:conf/eccv/ZhangLLWZF18} & RealSR\cite{cai2019toward}    \\
         		\specialrule{0em}{-8pt}{0pt}
         		\subfloat{\includegraphics[width=0.16\textwidth,height=0.12\textwidth]{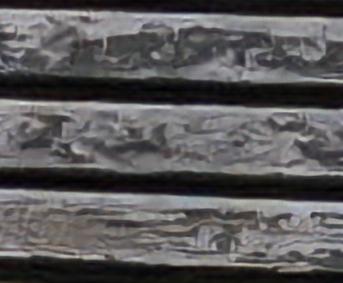}} 
				&
         		\subfloat{\includegraphics[width=0.16\textwidth,height=0.12\textwidth]{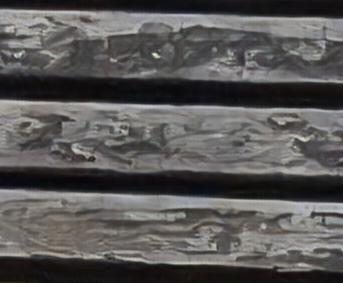}} 
         		&
         		\subfloat{\includegraphics[width=0.16\textwidth,height=0.12\textwidth]{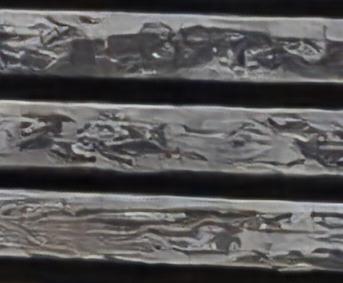}} 
         		&
         		\subfloat{\includegraphics[width=0.16\textwidth,height=0.12\textwidth]{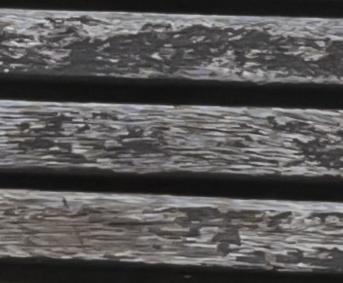}} \\
         		CinCGAN \cite{DBLP:conf/cvpr/YuanLZZDL18} & Ours-NoOver & Ours & Real HR Image \\  
			\end{tabular}
		\end{adjustbox}           
    \end{tabular}
    
    \begin{tabular}{cc}
    	\begin{adjustbox}{valign=t}
			\begin{tabular}{c}
			\specialrule{0em}{-8pt}{0pt}
			\subfloat{\includegraphics[width=0.23\textwidth,height=0.17\textwidth]{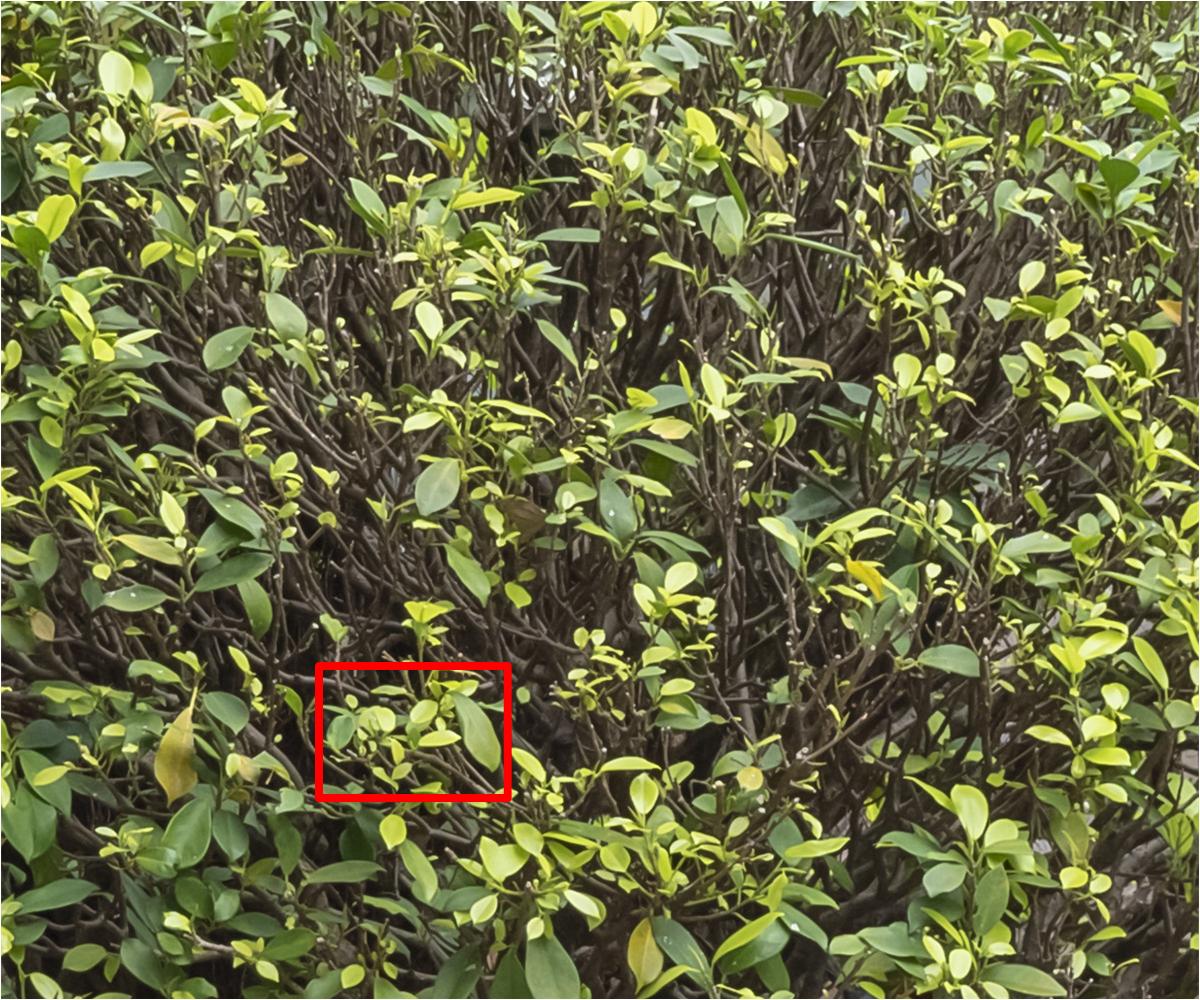}} \\
			rattan
			\end{tabular}
		\end{adjustbox}
		\begin{adjustbox}{valign=t}
			\begin{tabular}{c@{\hspace{1mm}}c@{\hspace{1mm}}c@{\hspace{1mm}}c}
			\specialrule{0em}{-8pt}{0pt}
				\subfloat{\includegraphics[width=0.16\textwidth,height=0.12\textwidth]{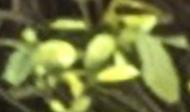}} 
				&
				\subfloat{\includegraphics[width=0.16\textwidth,height=0.12\textwidth]{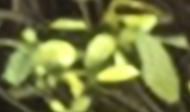}} 
				&
         		\subfloat{\includegraphics[width=0.16\textwidth,height=0.12\textwidth]{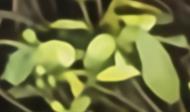}} 
         		&
				\subfloat{\includegraphics[width=0.16\textwidth,height=0.12\textwidth]{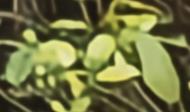}} 
         		\\
         		ESRGAN-Bic \cite{DBLP:conf/eccv/WangYWGLDQL18} & RCAN-Bic \cite{DBLP:conf/eccv/ZhangLLWZF18}  & RCAN-Real \cite{DBLP:conf/eccv/ZhangLLWZF18} & RealSR\cite{cai2019toward}    \\
         		\specialrule{0em}{-8pt}{0pt}
         		\subfloat{\includegraphics[width=0.16\textwidth,height=0.12\textwidth]{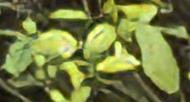}} 
				&
         		\subfloat{\includegraphics[width=0.16\textwidth,height=0.12\textwidth]{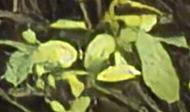}} 
         		&
         		\subfloat{\includegraphics[width=0.16\textwidth,height=0.12\textwidth]{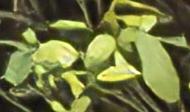}} 
         		&
         		\subfloat{\includegraphics[width=0.16\textwidth,height=0.12\textwidth]{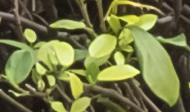}} \\
         		CinCGAN \cite{DBLP:conf/cvpr/YuanLZZDL18} & Ours-NoOver & Ours & Real HR Image \\  
			\end{tabular}
		\end{adjustbox}           
    \end{tabular}
    
    \begin{tabular}{cc}
    	\begin{adjustbox}{valign=t}
			\begin{tabular}{c}
			\specialrule{0em}{-8pt}{0pt}
				\subfloat{\includegraphics[width=0.23\textwidth,height=0.17\textwidth]{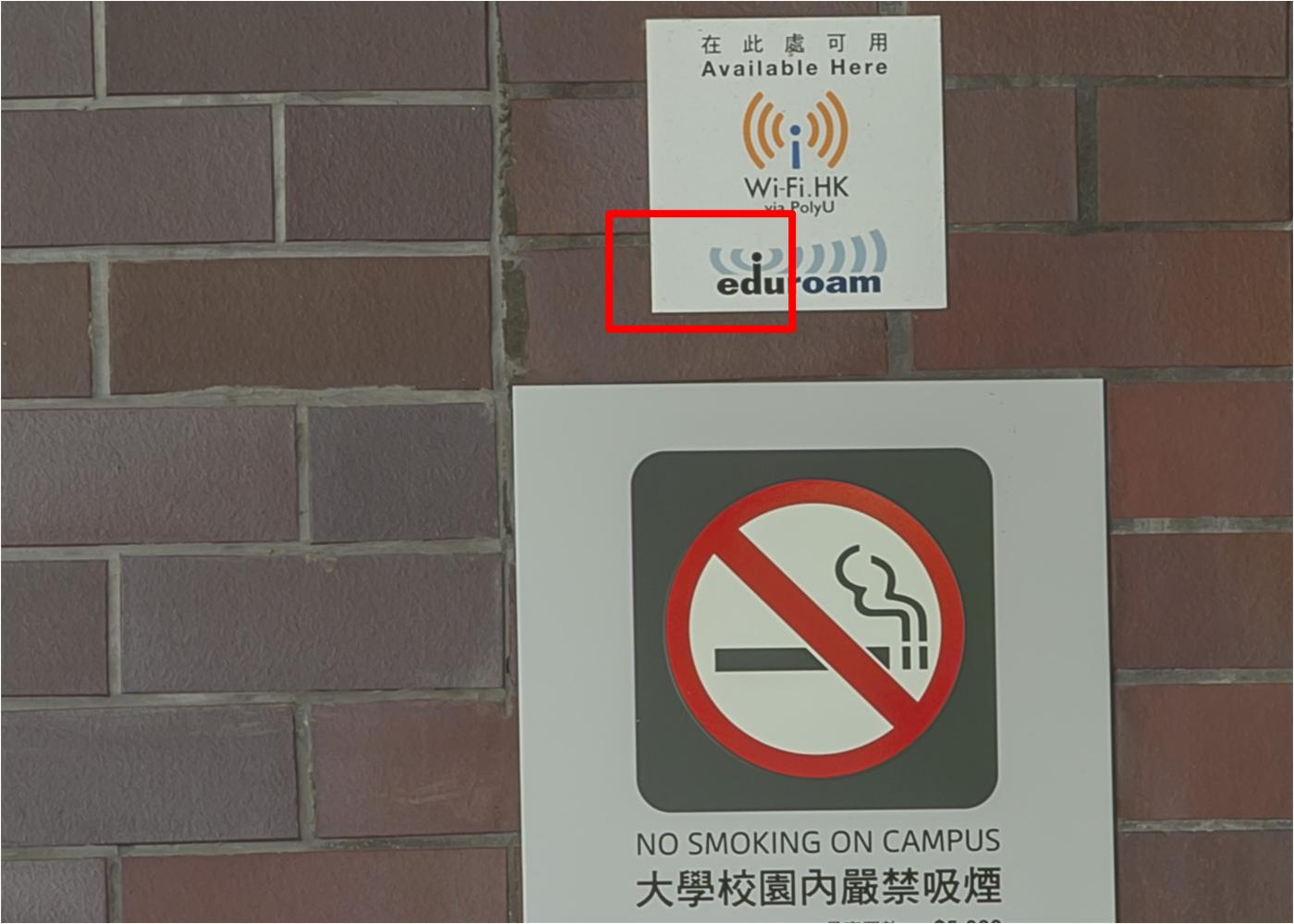}} \\
				smoke
			\end{tabular}
		\end{adjustbox}
		\begin{adjustbox}{valign=t}
			\begin{tabular}{c@{\hspace{1mm}}c@{\hspace{1mm}}c@{\hspace{1mm}}c}
			    \specialrule{0em}{-8pt}{0pt}
				\subfloat{\includegraphics[width=0.16\textwidth,height=0.12\textwidth]{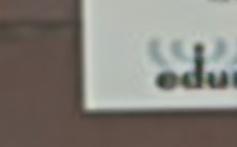}} 
				&
				\subfloat{\includegraphics[width=0.16\textwidth,height=0.12\textwidth]{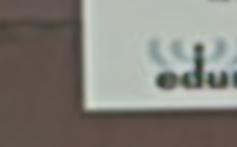}} 
				&
         		\subfloat{\includegraphics[width=0.16\textwidth,height=0.12\textwidth]{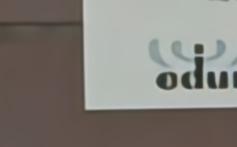}} 
         		&
         		\subfloat{\includegraphics[width=0.16\textwidth,height=0.12\textwidth]{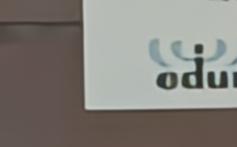}} 
         		\\
         		ESRGAN-Bic \cite{DBLP:conf/eccv/WangYWGLDQL18} & RCAN-Bic \cite{DBLP:conf/eccv/ZhangLLWZF18}    & RCAN-Real\cite{DBLP:conf/eccv/ZhangLLWZF18} & RealSR \cite{cai2019toward}    \\
         		\specialrule{0em}{-8pt}{0pt}
				\subfloat{\includegraphics[width=0.16\textwidth,height=0.12\textwidth]{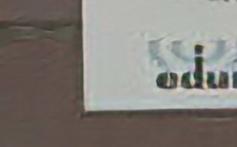}} 
         		&
         		\subfloat{\includegraphics[width=0.16\textwidth,height=0.12\textwidth]{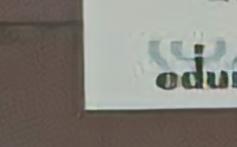}} 
         		&
         		\subfloat{\includegraphics[width=0.16\textwidth,height=0.12\textwidth]{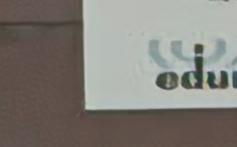}} 
         		&
         		\subfloat{\includegraphics[width=0.16\textwidth,height=0.12\textwidth]{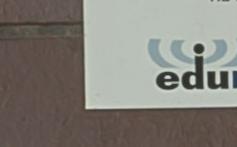}} \\
         		CinCGAN \cite{DBLP:conf/cvpr/YuanLZZDL18} & Ours-NoOver & Ours & Real HR Image \\  
			\end{tabular}
		\end{adjustbox}           
    \end{tabular}
    
    \begin{tabular}{cc}
    	\begin{adjustbox}{valign=t}
			\begin{tabular}{c}
			\specialrule{0em}{-8pt}{0pt}
				\subfloat{\includegraphics[width=0.23\textwidth,height=0.17\textwidth]{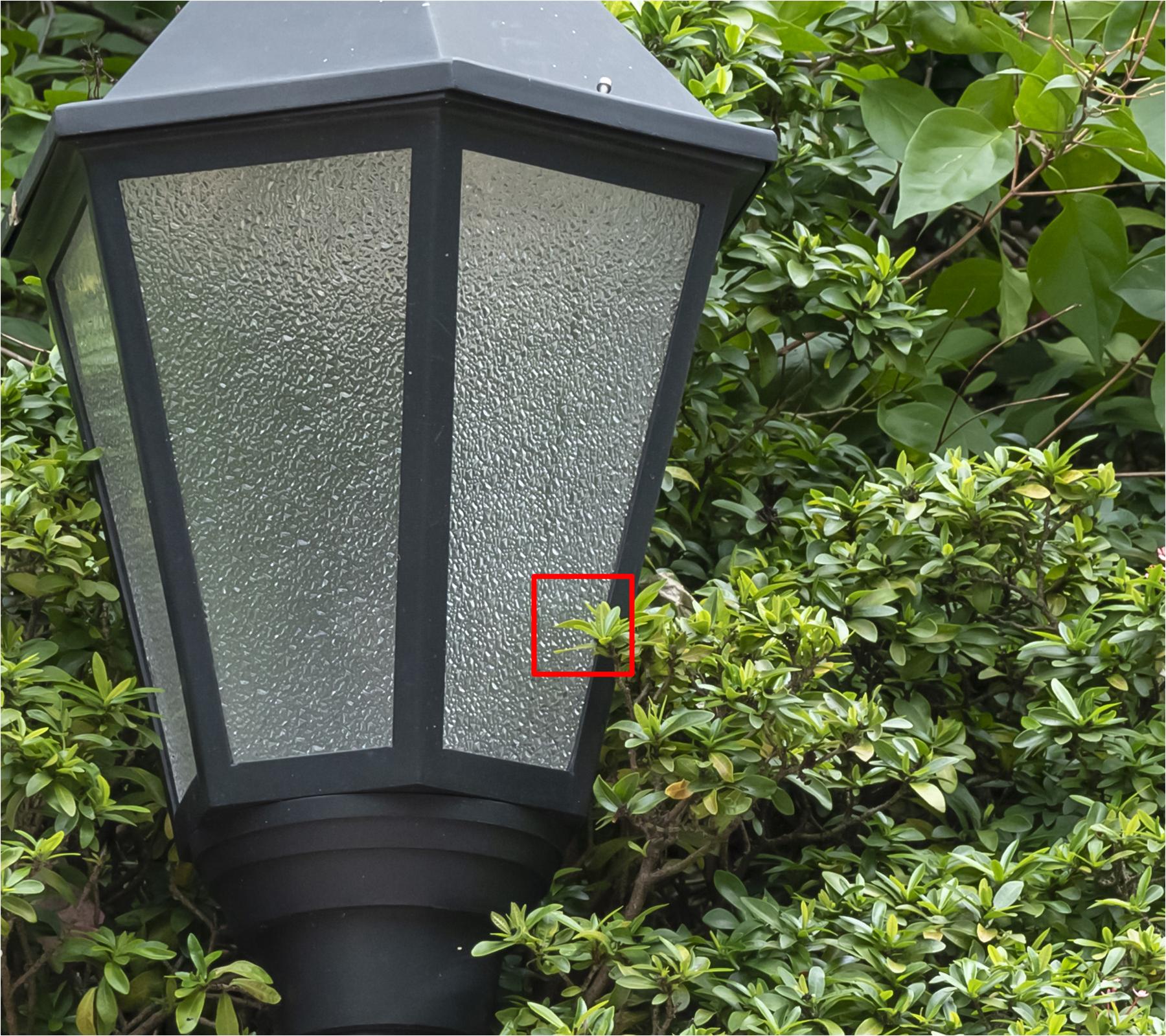}} \\
				light
			\end{tabular}
		\end{adjustbox}
		\begin{adjustbox}{valign=t}
			\begin{tabular}{c@{\hspace{1mm}}c@{\hspace{1mm}}c@{\hspace{1mm}}c}
			\specialrule{0em}{-8pt}{0pt}
				\subfloat{\includegraphics[width=0.16\textwidth,height=0.12\textwidth]{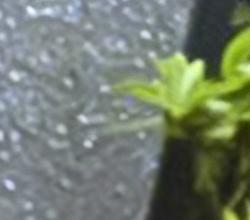}} 
				&
				\subfloat{\includegraphics[width=0.16\textwidth,height=0.12\textwidth]{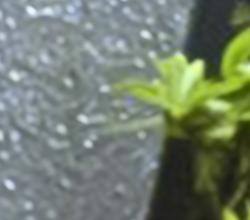}} 
				&
         		\subfloat{\includegraphics[width=0.16\textwidth,height=0.12\textwidth]{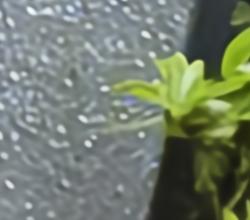}} 
         		&
				\subfloat{\includegraphics[width=0.16\textwidth,height=0.12\textwidth]{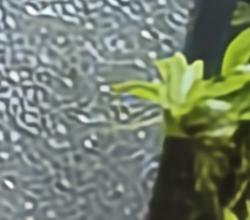}} 
         		\\
         		ESRGAN-Bic \cite{DBLP:conf/eccv/WangYWGLDQL18} & RCAN-Bic \cite{DBLP:conf/eccv/ZhangLLWZF18}  & RCAN-Real \cite{DBLP:conf/eccv/ZhangLLWZF18} & RealSR\cite{cai2019toward}    \\
         		\specialrule{0em}{-8pt}{0pt}
         		\subfloat{\includegraphics[width=0.16\textwidth,height=0.12\textwidth]{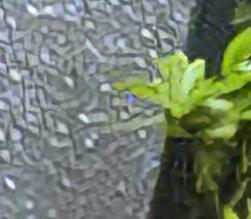}} 
				&
         		\subfloat{\includegraphics[width=0.16\textwidth,height=0.12\textwidth]{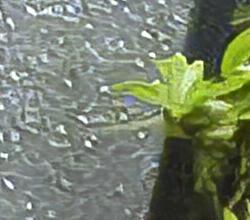}} 
         		&
         		\subfloat{\includegraphics[width=0.16\textwidth,height=0.12\textwidth]{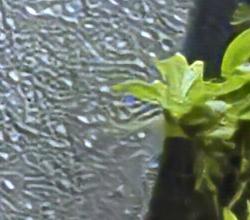}} 
         		&
         		\subfloat{\includegraphics[width=0.16\textwidth,height=0.12\textwidth]{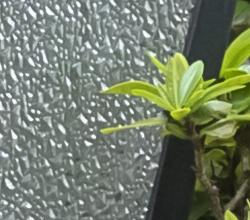}} \\
         		CinCGAN \cite{DBLP:conf/cvpr/YuanLZZDL18} & Ours-NoOver & Ours & Real HR Image \\  
			\end{tabular}
		\end{adjustbox}           
    \end{tabular}

    \vspace{-0.2cm}
    \caption{Visual comparisons for 4x SR on the testing set from RealSR dataset \cite{cai2019toward}. Pre-registration is applied to the training LR-HR image pairs. Our approach obtains results with better visual quality, such as the textures in ``bench'' and ``rattan'', character ``e'' in ``smoke'' and details of the ``light''.}
    \vspace{-0.5cm}
    \label{fig:visual_RealSR}
\end{figure*}

\begin{figure*}[]
\footnotesize
	\centering
    
    \begin{tabular}{cc}
    	\begin{adjustbox}{valign=t}
			\begin{tabular}{c}
			\subfloat{\includegraphics[width=0.23\textwidth,height=0.17\textwidth]{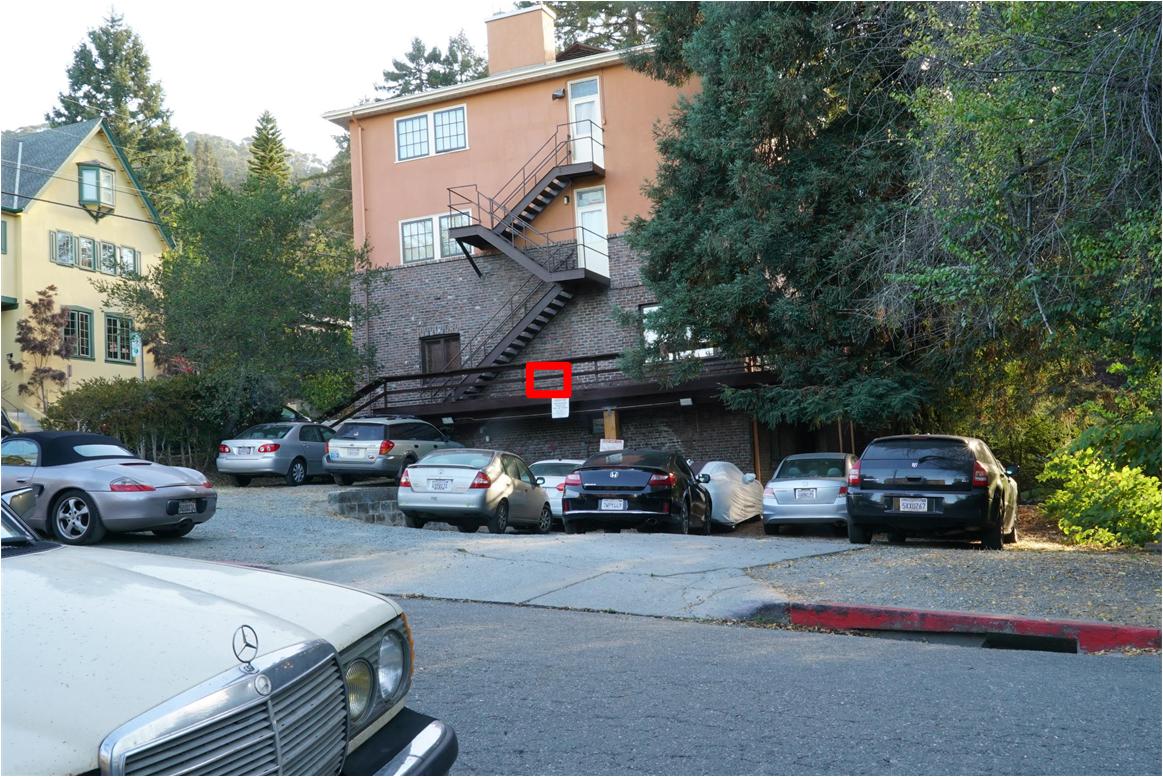}} \\
			brick
			\end{tabular}
		\end{adjustbox}
		\begin{adjustbox}{valign=t}
			\begin{tabular}{c@{\hspace{1mm}}c@{\hspace{1mm}}c@{\hspace{1mm}}c}
				\subfloat{\includegraphics[width=0.16\textwidth,height=0.12\textwidth]{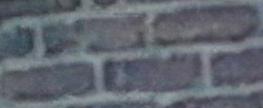}} 
				&
				\subfloat{\includegraphics[width=0.16\textwidth,height=0.12\textwidth]{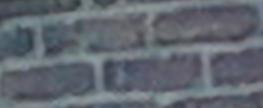}} 
				&
         		\subfloat{\includegraphics[width=0.16\textwidth,height=0.12\textwidth]{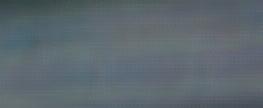}} 
         		&
				\subfloat{\includegraphics[width=0.16\textwidth,height=0.12\textwidth]{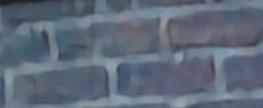}} 
         		\\
         		ESRGAN-Bic \cite{DBLP:conf/eccv/WangYWGLDQL18} & RCAN-Bic \cite{DBLP:conf/eccv/ZhangLLWZF18}  & RCAN-Real \cite{DBLP:conf/eccv/ZhangLLWZF18} & ZoomSR \cite{DBLP:conf/cvpr/ZhangCNK19}    \\
         		\specialrule{0em}{-8pt}{0pt}
         		\subfloat{\includegraphics[width=0.16\textwidth,height=0.12\textwidth]{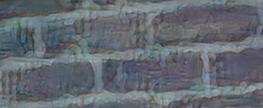}} 
				&
         		\subfloat{\includegraphics[width=0.16\textwidth,height=0.12\textwidth]{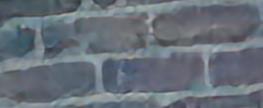}} 
         		&
         		\subfloat{\includegraphics[width=0.16\textwidth,height=0.12\textwidth]{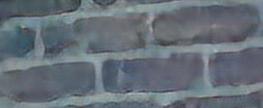}} 
         		&
         		\subfloat{\includegraphics[width=0.16\textwidth,height=0.12\textwidth]{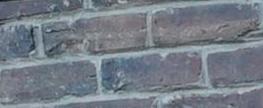}} \\
         		CinCGAN \cite{DBLP:conf/cvpr/YuanLZZDL18} & Ours-NoOver & Ours & Real HR Image \\  
			\end{tabular}
		\end{adjustbox}           
    \end{tabular}
    
    \begin{tabular}{cc}
    	\begin{adjustbox}{valign=t}
			\begin{tabular}{c}
			\specialrule{0em}{-8pt}{0pt}
			\subfloat{\includegraphics[width=0.23\textwidth,height=0.17\textwidth]{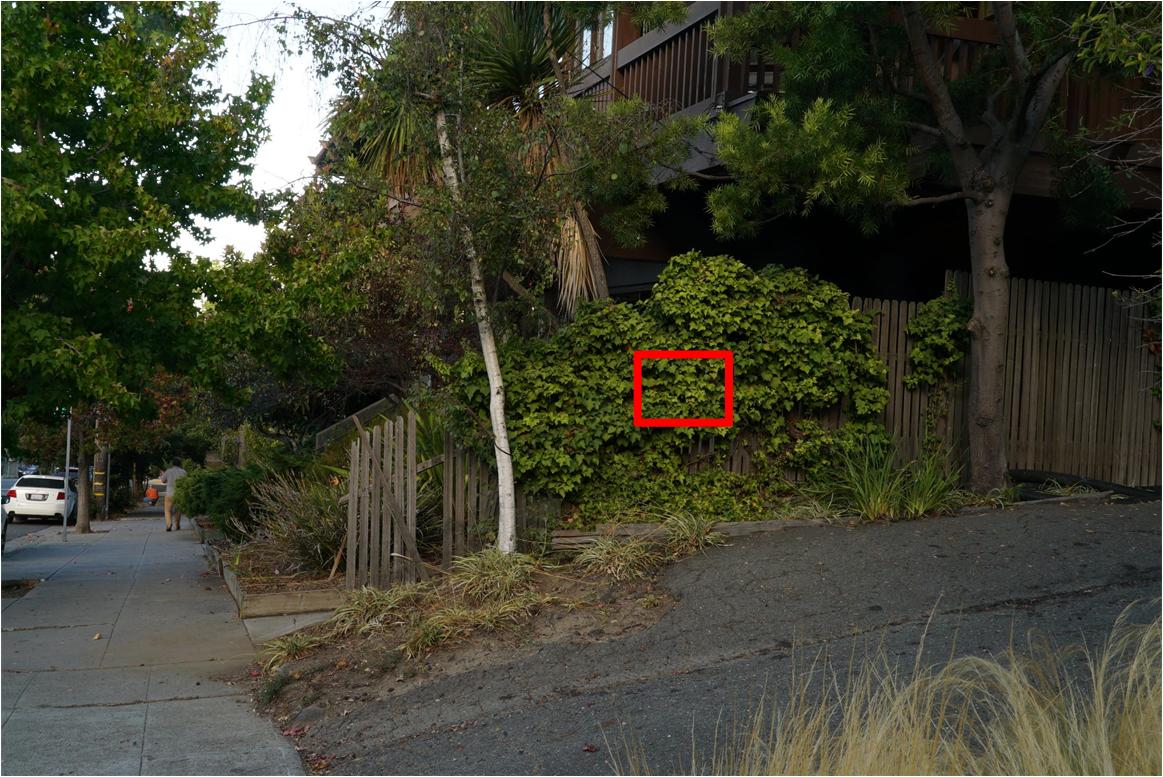}} \\
			leaf
			\end{tabular}
		\end{adjustbox}
		\begin{adjustbox}{valign=t}
			\begin{tabular}{c@{\hspace{1mm}}c@{\hspace{1mm}}c@{\hspace{1mm}}c}
			\specialrule{0em}{-8pt}{0pt}
				\subfloat{\includegraphics[width=0.16\textwidth,height=0.12\textwidth]{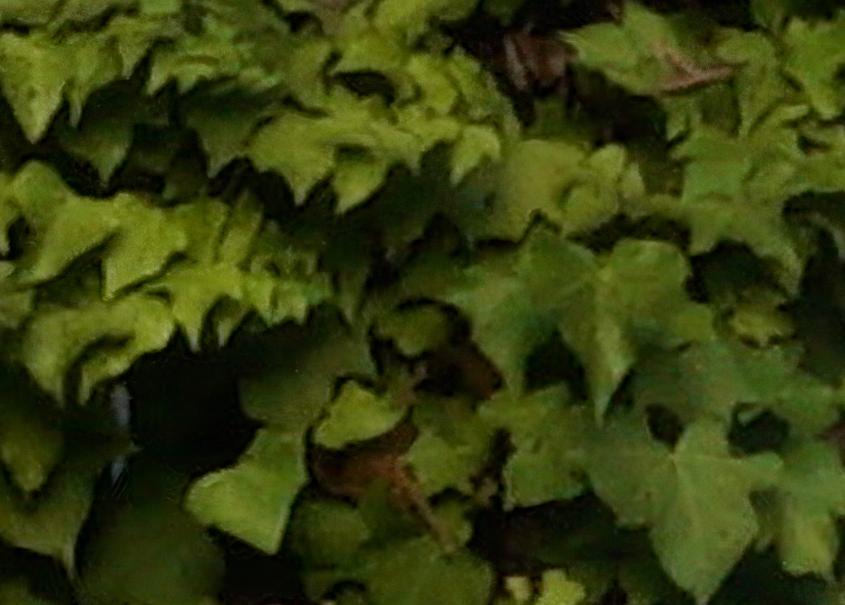}} 
				&
				\subfloat{\includegraphics[width=0.16\textwidth,height=0.12\textwidth]{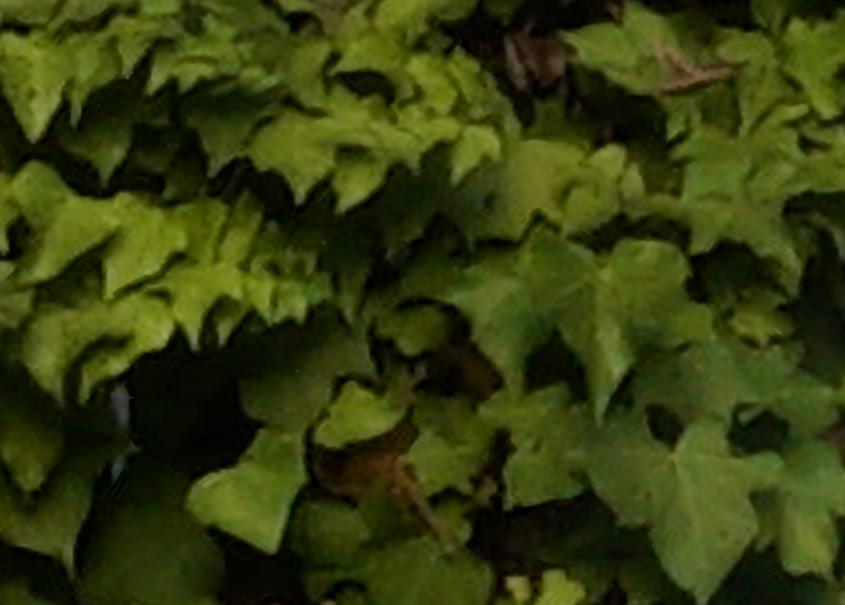}} 
				&
         		\subfloat{\includegraphics[width=0.16\textwidth,height=0.12\textwidth]{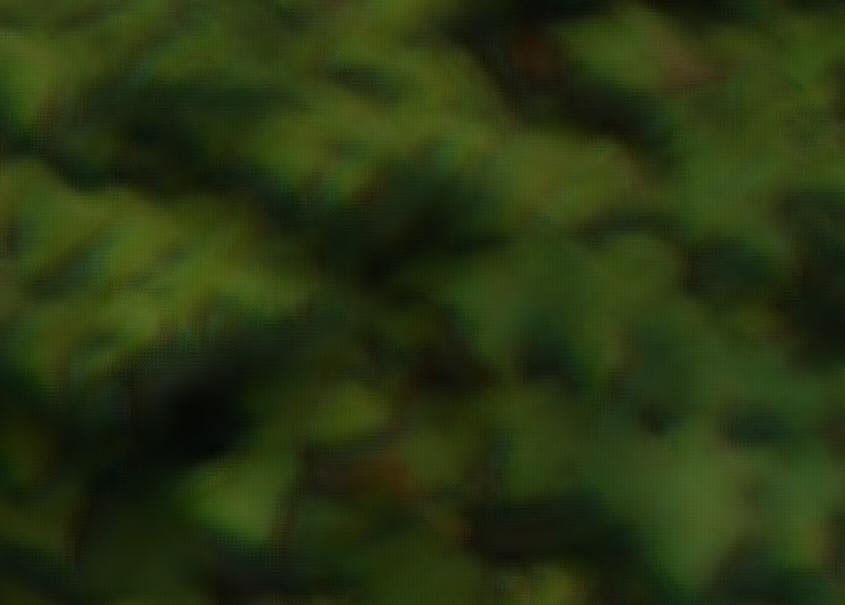}} 
         		&
				\subfloat{\includegraphics[width=0.16\textwidth,height=0.12\textwidth]{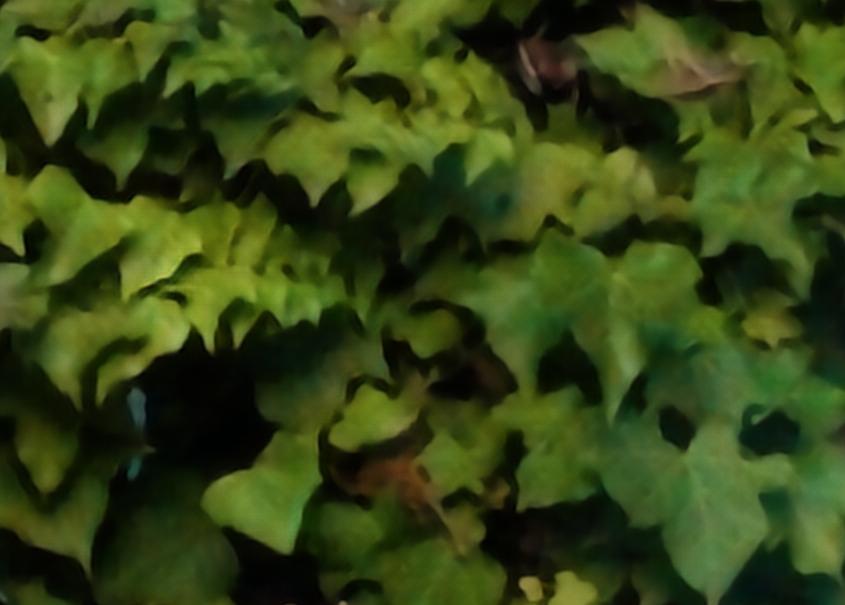}} 
         		\\
         		ESRGAN-Bic \cite{DBLP:conf/eccv/WangYWGLDQL18} & RCAN-Bic \cite{DBLP:conf/eccv/ZhangLLWZF18}  & RCAN-Real \cite{DBLP:conf/eccv/ZhangLLWZF18} & ZoomSR \cite{DBLP:conf/cvpr/ZhangCNK19}    \\
         		\specialrule{0em}{-8pt}{0pt}
         		\subfloat{\includegraphics[width=0.16\textwidth,height=0.12\textwidth]{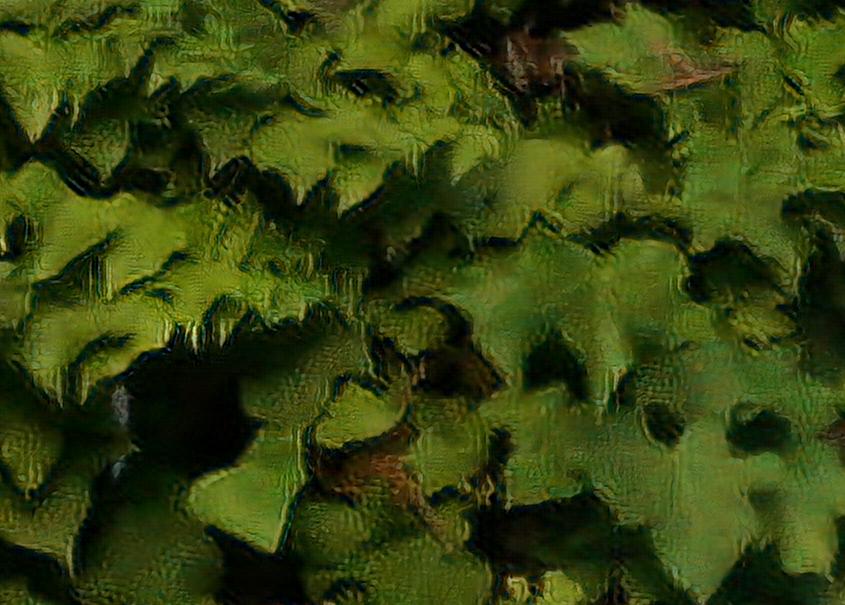}} 
				&
         		\subfloat{\includegraphics[width=0.16\textwidth,height=0.12\textwidth]{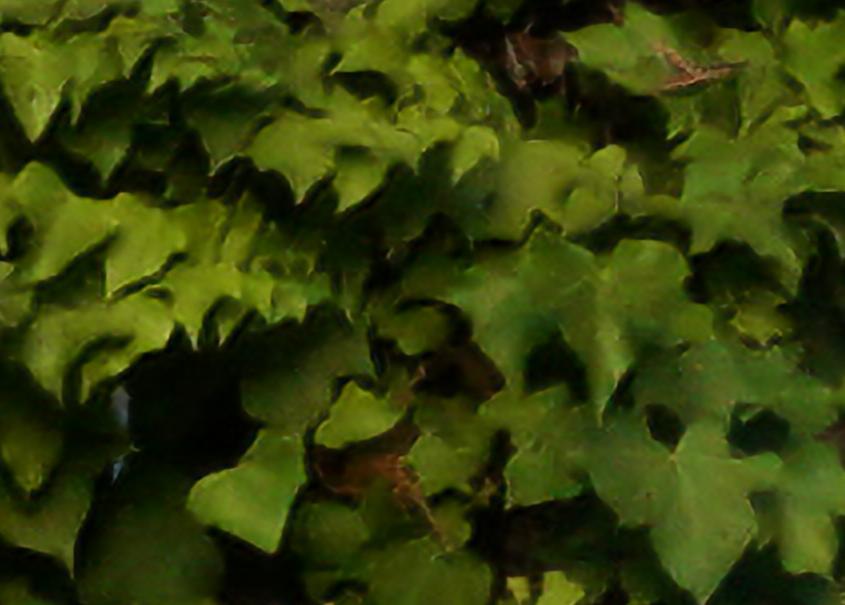}} 
         		&
         		\subfloat{\includegraphics[width=0.16\textwidth,height=0.12\textwidth]{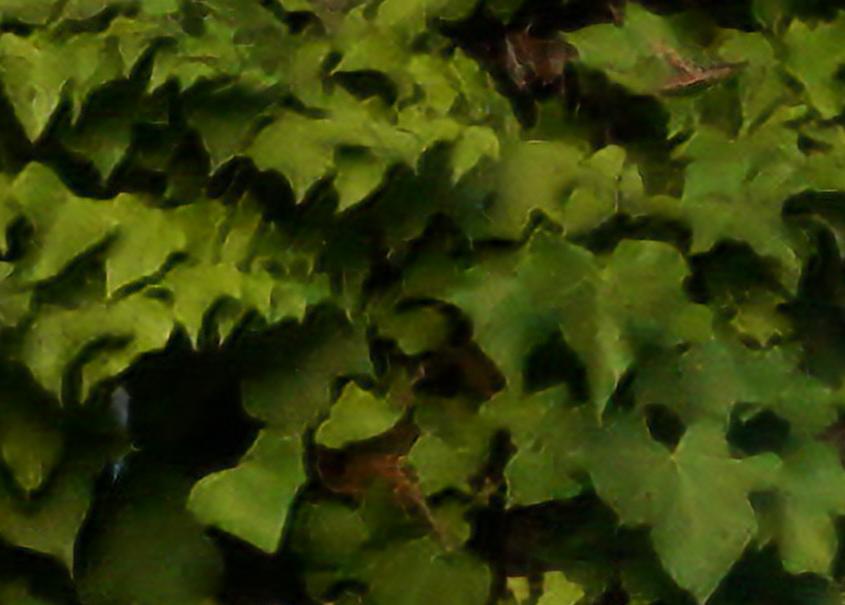}} 
         		&
         		\subfloat{\includegraphics[width=0.16\textwidth,height=0.12\textwidth]{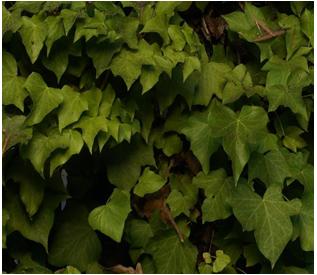}} \\
         		CinCGAN \cite{DBLP:conf/cvpr/YuanLZZDL18} & Ours-NoOver & Ours & Real HR Image \\  
			\end{tabular}
		\end{adjustbox}           
    \end{tabular}
    
    \begin{tabular}{cc}
    	\begin{adjustbox}{valign=t}
			\begin{tabular}{c}
			\specialrule{0em}{-8pt}{0pt}
			\subfloat{\includegraphics[width=0.23\textwidth,height=0.17\textwidth]{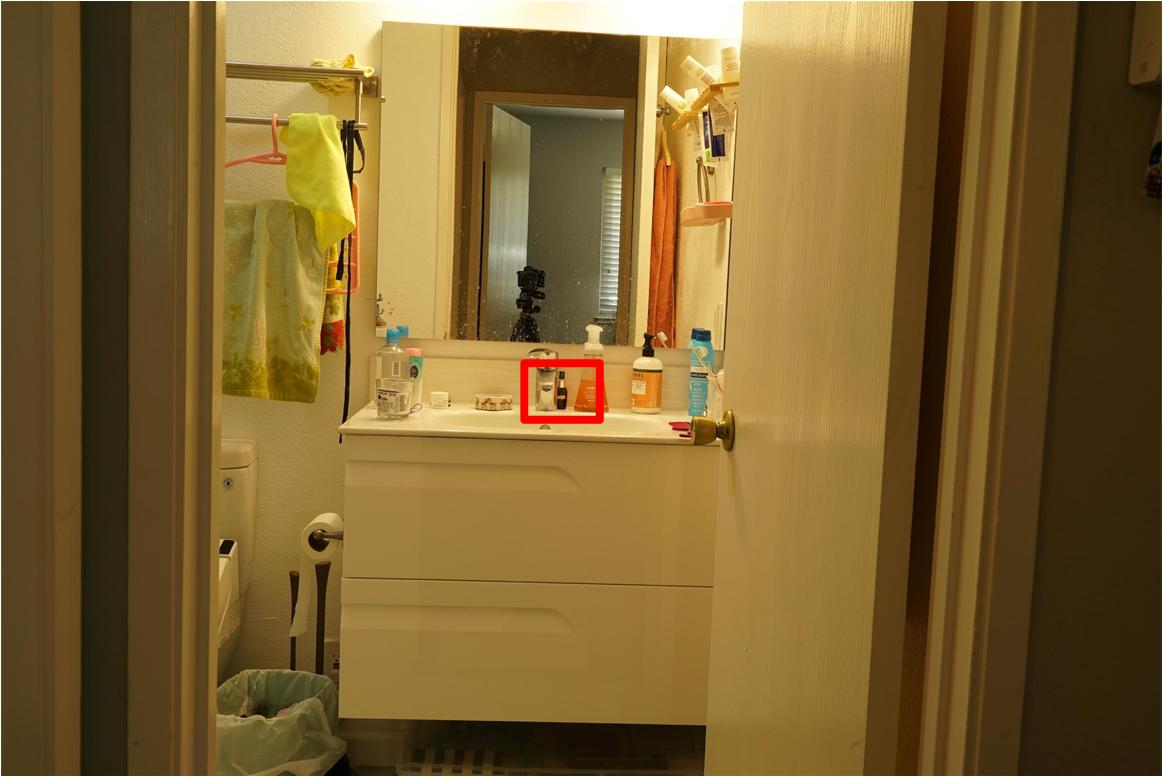}} \\
			bottle
			\end{tabular}
		\end{adjustbox}
		\begin{adjustbox}{valign=t}
			\begin{tabular}{c@{\hspace{1mm}}c@{\hspace{1mm}}c@{\hspace{1mm}}c}
			\specialrule{0em}{-8pt}{0pt}
				\subfloat{\includegraphics[width=0.16\textwidth,height=0.18\textwidth]{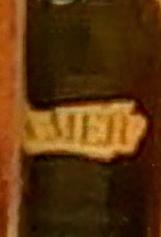}} 
				&
				\subfloat{\includegraphics[width=0.16\textwidth,height=0.18\textwidth]{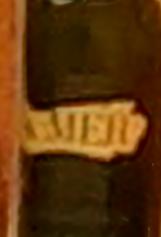}} 
				&
         		\subfloat{\includegraphics[width=0.16\textwidth,height=0.18\textwidth]{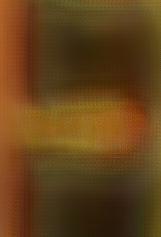}} 
         		&
				\subfloat{\includegraphics[width=0.16\textwidth,height=0.18\textwidth]{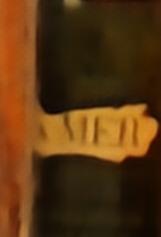}} 
         		\\
         		ESRGAN-Bic \cite{DBLP:conf/eccv/WangYWGLDQL18} & RCAN-Bic \cite{DBLP:conf/eccv/ZhangLLWZF18}  & RCAN-Real \cite{DBLP:conf/eccv/ZhangLLWZF18} & ZoomSR \cite{DBLP:conf/cvpr/ZhangCNK19}    \\
         		\specialrule{0em}{-8pt}{0pt}
         		\subfloat{\includegraphics[width=0.16\textwidth,height=0.18\textwidth]{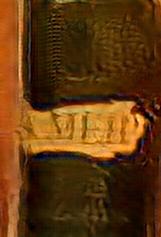}} 
				&
         		\subfloat{\includegraphics[width=0.16\textwidth,height=0.18\textwidth]{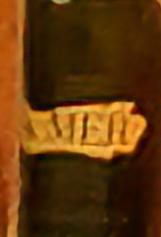}} 
         		&
         		\subfloat{\includegraphics[width=0.16\textwidth,height=0.18\textwidth]{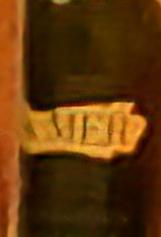}} 
         		&
         		\subfloat{\includegraphics[width=0.16\textwidth,height=0.18\textwidth]{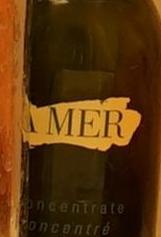}} \\
         		CinCGAN \cite{DBLP:conf/cvpr/YuanLZZDL18} & Ours-NoOver & Ours & Real HR Image \\  
			\end{tabular}
		\end{adjustbox}           
    \end{tabular}
    
    \vspace{-0.2cm}
    \caption{Visual comparisons for 4x SR on the testing set from SR-RGB dataset \cite{DBLP:conf/cvpr/ZhangCNK19}. Misalignment exists in LR-HR image pairs as no pre-registration is applied. The non-aligned real HR image is shown for reference. }
    \vspace{-0.5cm}
    \label{fig:visual_SR-RGB}
\end{figure*}

{\small
\bibliographystyle{ieee_fullname}
\bibliography{egbib}

\begin{thebibliography}{10}\itemsep=-1pt

\bibitem{DBLP:conf/eccv/BlauMTMZ18}
Yochai Blau, Roey Mechrez, Radu Timofte, Tomer Michaeli, and Lihi
  Zelnik{-}Manor.
\newblock The 2018 {PIRM} challenge on perceptual image super-resolution.
\newblock In {\em In European conference on computer vision Workshops (ECCVW)},
  pages 334--355, 2018.

\bibitem{DBLP:conf/cvpr/BlauM18}
Yochai Blau and Tomer Michaeli.
\newblock The perception-distortion tradeoff.
\newblock In {\em In {IEEE} Conference on Computer Vision and Pattern
  Recognition (CVPR)}, pages 6228--6237, 2018.

\bibitem{DBLP:journals/corr/BrunaSL15}
Joan Bruna, Pablo Sprechmann, and Yann LeCun.
\newblock Super-resolution with deep convolutional sufficient statistics.
\newblock In {\em International Conference on Learning Representations (ICLR)},
  2016.

\bibitem{DBLP:conf/eccv/BulatYT18}
Adrian Bulat, Jing Yang, and Georgios Tzimiropoulos.
\newblock To learn image super-resolution, use a {GAN} to learn how to do image
  degradation first.
\newblock In {\em In European conference on computer vision (ECCV)}, pages
  187--202, 2018.

\bibitem{cai2019toward}
Jianrui Cai, Hui Zeng, Hongwei Yong, Zisheng Cao, and Lei Zhang.
\newblock Toward real-world single image super-resolution: A new benchmark and
  a new model.
\newblock In {\em In IEEE International Conference on Computer Vision (ICCV)},
  2019.

\bibitem{DBLP:conf/cvpr/ChenXTZW19}
Chang Chen, Zhiwei Xiong, Xinmei Tian, Zheng{-}Jun Zha, and Feng Wu.
\newblock Camera lens super-resolution.
\newblock In {\em In {IEEE} Conference on Computer Vision and Pattern
  Recognition (CVPR)}, pages 1652--1660, 2019.

\bibitem{DBLP:journals/pami/DongLHT16}
Chao Dong, Chen~Change Loy, Kaiming He, and Xiaoou Tang.
\newblock Image super-resolution using deep convolutional networks.
\newblock {\em IEEE transactions on pattern analysis and machine intelligence},
  38(2):295--307, 2016.

\bibitem{gong2017motion}
Dong Gong, Jie Yang, Lingqiao Liu, Yanning Zhang, Ian Reid, Chunhua Shen, Anton
  Van Den~Hengel, and Qinfeng Shi.
\newblock From motion blur to motion flow: a deep learning solution for
  removing heterogeneous motion blur.
\newblock In {\em Proceedings of the IEEE Conference on Computer Vision and
  Pattern Recognition (CVPR)}, pages 2319--2328, 2017.

\bibitem{DBLP:conf/nips/GoodfellowPMXWOCB14}
Ian~J. Goodfellow, Jean Pouget{-}Abadie, Mehdi Mirza, Bing Xu, David
  Warde{-}Farley, Sherjil Ozair, Aaron~C. Courville, and Yoshua Bengio.
\newblock Generative adversarial nets.
\newblock In {\em Advances in neural information processing systems}, pages
  2672--2680, 2014.

\bibitem{DBLP:conf/cvpr/HarisSU18}
Muhammad Haris, Gregory Shakhnarovich, and Norimichi Ukita.
\newblock Deep back-projection networks for super-resolution.
\newblock In {\em In IEEE conference on computer vision and pattern recognition
  (CVPR)}, pages 1664--1673, 2018.

\bibitem{DBLP:conf/cvpr/HeZRS16}
Kaiming He, Xiangyu Zhang, Shaoqing Ren, and Jian Sun.
\newblock Deep residual learning for image recognition.
\newblock In {\em In {IEEE} Conference on Computer Vision and Pattern
  Recognition (CVPR)}, pages 770--778, 2016.

\bibitem{DBLP:conf/cvpr/IsolaZZE17}
Phillip Isola, Jun{-}Yan Zhu, Tinghui Zhou, and Alexei~A. Efros.
\newblock Image-to-image translation with conditional adversarial networks.
\newblock In {\em In {IEEE} Conference on Computer Vision and Pattern
  Recognition (CVPR)}, pages 5967--5976, 2017.

\bibitem{DBLP:conf/eccv/JohnsonAF16}
Justin Johnson, Alexandre Alahi, and Li Fei{-}Fei.
\newblock Perceptual losses for real-time style transfer and super-resolution.
\newblock In {\em In European conference on computer vision (ECCV)}, pages
  694--711, 2016.

\bibitem{DBLP:conf/iclr/Jolicoeur-Martineau19}
Alexia Jolicoeur{-}Martineau.
\newblock The relativistic discriminator: a key element missing from standard
  {GAN}.
\newblock In {\em In International Conference on Learning Representations
  (ICLR)}, 2019.

\bibitem{DBLP:conf/cvpr/LedigTHCCAATTWS17}
Christian Ledig, Lucas Theis, Ferenc Huszar, Jose Caballero, Andrew Cunningham,
  Alejandro Acosta, Andrew~P. Aitken, Alykhan Tejani, Johannes Totz, Zehan
  Wang, and Wenzhe Shi.
\newblock Photo-realistic single image super-resolution using a generative
  adversarial network.
\newblock In {\em In {IEEE} Conference on Computer Vision and Pattern
  Recognition (CVPR)}, pages 105--114, 2017.

\bibitem{DBLP:conf/eccv/LiW16}
Chuan Li and Michael Wand.
\newblock Precomputed real-time texture synthesis with markovian generative
  adversarial networks.
\newblock In {\em In European Conference on Computer Vision (ECCV)}, pages
  702--716, 2016.

\bibitem{DBLP:conf/cvpr/LimSKNL17}
Bee Lim, Sanghyun Son, Heewon Kim, Seungjun Nah, and Kyoung~Mu Lee.
\newblock Enhanced deep residual networks for single image super-resolution.
\newblock In {\em In {IEEE} Conference on Computer Vision and Pattern
  Recognition Workshops (CVPRW)}, pages 1132--1140, 2017.

\bibitem{DBLP:journals/corr/abs-1909-09629}
Andreas Lugmayr, Martin Danelljan, and Radu Timofte.
\newblock Unsupervised learning for real-world super-resolution.
\newblock {\em arXiv preprint arXiv:1909.09629}, 2019.

\bibitem{shen2019regularizing}
Tong Shen, Dong Gong, Wei Zhang, Chunhua Shen, and Tao Mei.
\newblock Regularizing proxies with multi-adversarial training for unsupervised
  domain-adaptive semantic segmentation.
\newblock {\em arXiv preprint arXiv:1907.12282}, 2019.

\bibitem{DBLP:conf/cvpr/TimofteGWG18}
Radu Timofte, Shuhang Gu, Jiqing Wu, and Luc~Van Gool.
\newblock {NTIRE} 2018 challenge on single image super-resolution: Methods and
  results.
\newblock In {\em In {IEEE} Conference on Computer Vision and Pattern
  Recognition Workshops (CVPRW)}, pages 852--863, 2018.

\bibitem{DBLP:conf/iccv/0001LLG17}
Tong Tong, Gen Li, Xiejie Liu, and Qinquan Gao.
\newblock Image super-resolution using dense skip connections.
\newblock In {\em In {IEEE} International Conference on Computer Vision
  (ICCV)}, pages 4809--4817, 2017.

\bibitem{DBLP:conf/eccv/WangYWGLDQL18}
Xintao Wang, Ke Yu, Shixiang Wu, Jinjin Gu, Yihao Liu, Chao Dong, Yu Qiao, and
  Chen~Change Loy.
\newblock {ESRGAN:} enhanced super-resolution generative adversarial networks.
\newblock In {\em In European Conference on Computer Vision (ECCV)}, pages
  63--79, 2018.

\bibitem{DBLP:journals/tip/WangBSS04}
Zhou Wang, Alan~C. Bovik, Hamid~R. Sheikh, and Eero~P. Simoncelli.
\newblock Image quality assessment: from error visibility to structural
  similarity.
\newblock {\em IEEE transactions on image processing}, 13(4):600--612, 2004.

\bibitem{yang2018seeing}
Jie Yang, Dong Gong, Lingqiao Liu, and Qinfeng Shi.
\newblock Seeing deeply and bidirectionally: A deep learning approach for
  single image reflection removal.
\newblock In {\em Proceedings of the European Conference on Computer Vision
  (ECCV)}, pages 654--669, 2018.

\bibitem{DBLP:journals/tip/YangWHM10}
Jianchao Yang, John Wright, Thomas~S. Huang, and Yi Ma.
\newblock Image super-resolution via sparse representation.
\newblock {\em IEEE transactions on image processing}, 19(11):2861--2873, 2010.

\bibitem{DBLP:conf/iccv/YiZTG17}
Zili Yi, Hao~(Richard) Zhang, Ping Tan, and Minglun Gong.
\newblock Dualgan: Unsupervised dual learning for image-to-image translation.
\newblock In {\em In {IEEE} International Conference on Computer Vision
  (ICCV)}, pages 2868--2876, 2017.

\bibitem{DBLP:conf/cvpr/YuanLZZDL18}
Yuan Yuan, Siyuan Liu, Jiawei Zhang, Yongbing Zhang, Chao Dong, and Liang Lin.
\newblock Unsupervised image super-resolution using cycle-in-cycle generative
  adversarial networks.
\newblock In {\em In {IEEE} Conference on Computer Vision and Pattern
  Recognition Workshops (CVPRW)}, pages 701--710, 2018.

\bibitem{zhang2017beyond}
Kai Zhang, Wangmeng Zuo, Yunjin Chen, Deyu Meng, and Lei Zhang.
\newblock Beyond a gaussian denoiser: Residual learning of deep cnn for image
  denoising.
\newblock {\em IEEE Transactions on Image Processing}, 26(7):3142--3155, 2017.

\bibitem{DBLP:conf/cvpr/ZhangZ018}
Kai Zhang, Wangmeng Zuo, and Lei Zhang.
\newblock Learning a single convolutional super-resolution network for multiple
  degradations.
\newblock In {\em In {IEEE} Conference on Computer Vision and Pattern
  Recognition (CVPR)}, pages 3262--3271, 2018.

\bibitem{DBLP:conf/cvpr/ZhangIESW18}
Richard Zhang, Phillip Isola, Alexei~A. Efros, Eli Shechtman, and Oliver Wang.
\newblock The unreasonable effectiveness of deep features as a perceptual
  metric.
\newblock In {\em In {IEEE} Conference on Computer Vision and Pattern
  Recognition (CVPR)}, pages 586--595, 2018.

\bibitem{DBLP:conf/cvpr/ZhangCNK19}
Xuaner Zhang, Qifeng Chen, Ren Ng, and Vladlen Koltun.
\newblock Zoom to learn, learn to zoom.
\newblock In {\em In {IEEE} Conference on Computer Vision and Pattern
  Recognition (CVPR)}, pages 3762--3770, 2019.

\bibitem{DBLP:conf/eccv/ZhangLLWZF18}
Yulun Zhang, Kunpeng Li, Kai Li, Lichen Wang, Bineng Zhong, and Yun Fu.
\newblock Image super-resolution using very deep residual channel attention
  networks.
\newblock In {\em In European Conference on Computer Vision (ECCV)}, pages
  294--310, 2018.

\bibitem{DBLP:conf/iclr/ZhangLLZF19}
Yulun Zhang, Kunpeng Li, Kai Li, Bineng Zhong, and Yun Fu.
\newblock Residual non-local attention networks for image restoration.
\newblock In {\em In International Conference on Learning Representations
  (ICLR)}, 2019.

\bibitem{DBLP:conf/cvpr/ZhangTKZ018}
Yulun Zhang, Yapeng Tian, Yu Kong, Bineng Zhong, and Yun Fu.
\newblock Residual dense network for image super-resolution.
\newblock In {\em In {IEEE} Conference on Computer Vision and Pattern
  Recognition (CVPR)}, pages 2472--2481, 2018.

\bibitem{DBLP:conf/iccv/ZhuPIE17}
Jun{-}Yan Zhu, Taesung Park, Phillip Isola, and Alexei~A. Efros.
\newblock Unpaired image-to-image translation using cycle-consistent
  adversarial networks.
\newblock In {\em In {IEEE} International Conference on Computer Vision
  (ICCV)}, pages 2242--2251, 2017.

\end{thebibliography}
}

\end{document}